\documentclass[12pt]{article}
\usepackage{epsf}
\usepackage{amsfonts}
\usepackage{graphicx}
\usepackage{color}
\usepackage{psfrag}

\newcommand{\bmat}{\left(\begin{array}}
\newcommand{\emat}{\end{array}\right)}

\def\yzero{\smash{\hbox{$y\kern-4pt\raise1pt\hbox{${}^\circ$}$}}}

\def\g{\gamma}

\def\beq{\begin{equation}}
\def\eeq{\end{equation}}
\def\beqa{\begin{eqnarray}}
\def\eeqa{\end{eqnarray}}

\def\-{\hphantom{-}}

\def\s2{\frac{1}{\sqrt2}}

\def\Tr{{\rm Tr \,}}

\def\IF{\relax{\rm I\kern-.18em F}}
\def\II{\relax{\rm I\kern-.18em I}}
\def\IP{\relax{\rm I\kern-.18em P}}
\def\IC{\relax\hbox{\kern.25em$\inbar\kern-.3em{\rm C}$}}
\def\IR{\relax{\rm I\kern-.18em R}}

\def\cc{{\cal C}}

\def\cn{{\cal N}}
\def\cl{{\cal L}}
\def\cam{{\cal M}}

\def\Dsl{\,\raise.15ex\hbox{/}\mkern-13.5mu D} 
\def\IZ{Z\kern-.4em  Z}

\def\im{{\rm Im}\,}   
\def\re{{\rm Re}\,}

\catcode`\@=11   
\newdimen\@rotdimen
\newbox\@rotbox  

\def\@vspec#1{\special{ps:#1}}
\def\@rotstart#1{\@vspec{gsave currentpoint currentpoint translate
   #1 neg exch neg exch translate}}
\def\@rotfinish{\@vspec{currentpoint grestore moveto}}
\def\@rotr#1{\@rotdimen=\ht#1\advance\@rotdimen by\dp#1%
   \hbox to\@rotdimen{\hskip\ht#1\vbox to\wd#1{\@rotstart{90 rotate}%
   \box#1\vss}\hss}\@rotfinish}
\def\@rotl#1{\@rotdimen=\ht#1\advance\@rotdimen by\dp#1%
   \hbox to\@rotdimen{\vbox to\wd#1{\vskip\wd#1\@rotstart{270 rotate}%
   \box#1\vss}\hss}\@rotfinish}%
\def\@rotu#1{\@rotdimen=\ht#1\advance\@rotdimen by\dp#1%
   \hbox to\wd#1{\hskip\wd#1\vbox to\@rotdimen{\vskip\@rotdimen
   \@rotstart{-1 dup scale}\box#1\vss}\hss}\@rotfinish}%
\def\@rotf#1{\hbox to\wd#1{\hskip\wd#1\@rotstart{-1 1 scale}%
   \box#1\hss}\@rotfinish}%
\def\rotate{\@ifnextchar[{\@rotate}{\@rotate[l]}}
\def\@rotate[#1]#2{\setbox\@rotbox=\hbox{#2}\@nameuse{@rot#1}\@rotbox}

\catcode`\@=12

\topmargin
-1.5cm
\textwidth
15.5cm
\textheight
23.5cm
\oddsidemargin
0.7cm
\evensidemargin
1.2cm

\begin{document}

\makeatletter
\@addtoreset{equation}{section}
\makeatother
\renewcommand{\theequation}{\thesection.\arabic{equation}}
\pagestyle{empty}
\rightline{ IFT-UAM/CSIC-07-14}
\rightline{ CERN-PH-TH/2007-063}

\vspace{0.1cm}
\begin{center}
\LARGE{Supersymmetry breaking metastable vacua \\ in runaway quiver gauge theories \\[12mm]}
\large{I. Garc\'{\i}a-Etxebarria, F. Saad, A.M.Uranga\\[3mm]}
\footnotesize{PH-TH Division, CERN 
CH-1211 Geneva 23, Switzerland\\
 and \\
Instituto de F\'{\i}sica Te\'orica  C-XVI,\\[-0.3em]
Universidad Aut\'onoma de Madrid,
Cantoblanco, 28049 Madrid, Spain\\[2mm] }
\small{\bf Abstract} \\[5mm]
\end{center}
\begin{center}
\begin{minipage}[h]{16.0cm}
  In this paper we consider quiver gauge theories with fractional
  branes whose infrared dynamics removes the classical supersymmetric
  vacua (DSB branes). We show that addition of flavors to these
  theories (via additional non-compact branes) leads to local
  meta-stable supersymmetry breaking minima, closely related to those
  of SQCD with massive flavors. We simplify the study of the one-loop
  lifting of the accidental classical flat directions by direct
  computation of the pseudomoduli masses via Feynman diagrams. This
  new approach allows to obtain analytic results for all these
  theories. This work extends the results for the $dP_1$ theory in
  hep-th/0607218. The new approach allows to generalize the
  computation to general examples of DSB branes, and for arbitrary
  values of the superpotential couplings.
\end{minipage}
\end{center}
\newpage
\setcounter{page}{1}
\pagestyle{plain}
\renewcommand{\thefootnote}{\arabic{footnote}}
\setcounter{footnote}{0}


\section{Introduction}

Systems of D-branes at singularities provide a very interesting setup
to realize and study diverse non-perturbative gauge dynamics phenomena
in string theory. In the context of $\mathcal{N}=1$ supersymmetric
gauge field theories, systems of D3-branes at Calabi-Yau singularities
lead to interesting families of tractable 4d strongly coupled
conformal field theories, which extend the AdS/CFT correspondence
\cite{Maldacena:1997re,Gubser:1998bc,Witten:1998qj} to theories with
reduced (super)symmetry
\cite{Kachru:1998ys,Klebanov:1998hh,Morrison:1998cs} and enable
non-trivial precision tests of the correspondence (see for instance
\cite{Bertolini:2004xf,Benvenuti:2004dy}).  Addition of fractional
branes leads to families of non-conformal gauge theories, with
intricate RG flows involving cascades of Seiberg dualities
\cite{Klebanov:2000hb,Franco:2003ja,Franco:2004jz,Franco:2005fd,Herzog:2004tr},
and strong dynamics effects in the infrared.

For instance, fractional branes associated to complex deformations of
the singular geometry (denoted deformation fractional branes in
\cite{Franco:2005fd}), correspond to supersymmetric confinement of one
or several gauge factors in the gauge theory
\cite{Klebanov:2000hb,Franco:2005fd}. The generic case of fractional
branes associated to obstructed complex deformations (denoted DSB
branes in \cite{Franco:2005fd}), corresponds to gauge theories
developing a non-perturbative Affleck-Dine-Seiberg superpotential,
which removes the classical supersymmetric vacua
\cite{Berenstein:2005xa,Franco:2005zu,Bertolini:2005di}. As shown in
\cite{Franco:2005zu} (see also
\cite{Intriligator:2005aw,Brini:2006ej}), assuming canonical Kahler
potential leads to a runaway potential for the theory, along a
baryonic direction. A natural suggestion to stop this runaway has been
proposed for the particular example of the $dP_1$ theory (the theory
on fractional branes at the complex cone over $dP_1$) in
\cite{Franco:2006es}. It was shown that, upon the addition of
D7-branes to the configuration (which introduce massive flavors), the
theory develops a meta-stable minimum (closely related to the
Intriligator-Seiberg-Shih (ISS) model \cite{Intriligator:2006dd}),
parametrically long-lived against decay to the runaway regime (see
\cite{Florea:2006si} for an alternative suggestion to
stop the runaway, in compact models).

In this paper we show that the appearance of meta-stable minima in
gauge theories on DSB fractional branes, in the presence of additional
massless flavors, is much more general (and possibly valid in full
generality). We use the tools of \cite{Franco:2005zu} to introduce
D7-branes on general toric singularities, and give masses to the
corresponding flavors. Since quiver gauge theories are rather
involved, we develop new techniques to efficiently analyze the
one-loop stability of the meta-stable minima, via the direct
computation of Feynman diagrams. These tools can be used to argue
that the results plausibly hold for general systems of DSB fractional
branes at toric singularities. It is very satisfactory to verify the
correspondence between the existence of meta-stable vacua and the
geometric property of having obstructed complex deformations.

The present work thus enlarges the class of string models realizing
dynamical supersymmetry breaking in meta-stable vacua (see
\cite{Ooguri:2006bg,Argurio:2006ny,Franco:2006ht,Bena:2006rg,Argurio:2007qk}
for other proposed realizations, and
\cite{Lykken:1998ec,Wijnholt:2007vn,Antebi:2007xw} for models of
dynamical supersymmetry breaking in orientifold theories). Although we
will not discuss it in the present paper, these results can be applied
to the construction of models of gauge mediation in string theory as
in \cite{Garcia-Etxebarria:2006rw} (based on the additional tools in
\cite{Garcia-Etxebarria:2006aq}), in analogy with
\cite{Diaconescu:2005pc}. This is another motivation for the present
work.

The paper is organized as follows. In Section \ref{issrevisited} we
review the ISS model, evaluating one-loop pseudomoduli masses directly
in terms of Feynman diagrams. In Section \ref{dsb} we study the
theory of DSB branes at the $dP_1$ and $dP_2$ singularities upon the
addition of flavors, and we find that metastable vacua exist for these
theories. In Section \ref{generalcase} we extend this analysis to the
general case of DSB branes at toric singularities with massive
flavors, and we illustrate the results by showing the existence of
metastable vacua for DSB branes at some well known families of toric
singularities. Finally, the Appendix provides some technical details
that we have omitted from the main text in order to improve the
legibility.

\section{The ISS model revisited}
\label{issrevisited}

In this Section we review the ISS meta-stable minima in SQCD, and
propose that the analysis of the relevant piece of the one-loop
potential (the quadratic terms around the maximal symmetry point) is
most simply carried out by direct evaluation of Feynman diagrams.
This new tool will be most useful in the study of the more involved
examples of quiver gauge theories.

\subsection{The ISS metastable minimum}
\label{secISS}

The ISS model \cite{Intriligator:2006dd} (see also  \cite{Intriligator:2007cp} for a review of these and other models) is given by $\cn=1$ $SU(N_c)$
theory with $N_f$ flavors, with small masses 
\begin{equation}
\label{eq:iss-W-electric}
W_\textrm{electric} = m \Tr \phi \tilde\phi,
\end{equation}
where $\phi$ and $\tilde\phi$ are the quarks of the theory. The number of colors and flavors are chosen so as to be in the free magnetic phase:
\begin{equation}
N_c+1 \leq N_f < \frac{3}{2} N_c.
\end{equation}
This condition guarantees that the Seiberg dual is infrared free. This
Seiberg dual is the $SU(N)$ theory (with $N=N_f-N_c$) with $N_f$ flavors of dual
quarks $q$ and $\tilde q$ and the meson $M$. The dual superpotential
is given by rewriting (\ref{eq:iss-W-electric}) in terms of the mesons
and adding the usual coupling between the meson and the dual quarks:
\begin{equation}
\label{eq:iss-W-magnetic}
W_\textrm{magnetic}\, =\, h\, (\Tr \tilde q M q - \mu^2 \Tr M),
\end{equation}
where $h$ and $\mu$ can be expressed in terms of the parameters $m$
and $\Lambda$, and some (unknown) information about the dual K\"ahler
metric\footnote{The exact expressions can be found in (5.7) in
  \cite{Intriligator:2006dd}, but we will not need them for our
  analysis. We just take all masses in the electric description
 to be small enough for the analysis of the metastable vacuum to be
  reliable.}.  It was also argued in \cite{Intriligator:2006dd} that
it is possible to study the supersymmetry breaking minimum in the
origin of (dual) field space without taking into account the gauge
dynamics (their main effect in this discussion consists of restoring
supersymmetry dynamically far in field space). In the following we
will assume that this is always the case, and we will forget completely
about the gauge dynamics of the dual.

Once we forget about gauge dynamics, studying the vacua of the dual
theory becomes a matter of solving the F-term equations coming from
the superpotential (\ref{eq:iss-W-magnetic}). The mesonic F-term
equation reads:
\begin{equation}
-\overline F_{M_{ij}} = h \tilde q^i \cdot q^j - h \mu^2 \delta^{ij} = 0,
\end{equation}
where $i$ and $j$ are flavor indices and the dot denotes color
contraction. This has no solution, since the identity
matrix $\delta^{ij}$ has rank $N_f$ while $\tilde q^i \cdot q^j$ has
rank $N=N_f-N_c$. Thus this theory breaks supersymmetry spontaneously at tree level. This mechanism for F-term supersymmetry breaking is called the {\em rank condition}.

The classical scalar potential has a continuous set of minima, but the
one-loop potential lifts all of the non-Goldstone directions, which
are usually called pseudomoduli. The usual approach to study the
one-loop stabilization is the computation of the complete one-loop
effective potential over all pseudomoduli space via the
Coleman-Weinberg formula \cite{Coleman:1973jx}:
\begin{equation}
V = \frac{1}{64\pi^2}\Tr\left(\cam_B^4\log\frac{\cam_B^2}{\Lambda^2} -
\cam_F^4\log\frac{\cam_F^2}{\Lambda^2}\right).
\end{equation}
This approach has the advantage that it allows the determination
of the one-loop minimum, without {\em a priori} information about its
location, and moreover it provides the full potential around it,
including higher terms. However, it has the disadvantage of requiring
the diagonalization of the mass matrix, which very often does not
admit a closed expression, e.g. for the theories we are interested in.

In fact, we would like to point out that to determine the existence of
a meta-stable minimum there exists a computationally much simpler
approach. In our situation, we have a good ansatz for the location of
the one-loop minimum, and are interested just in the one-loop
pseudomoduli masses around such point. This information can be
directly obtained by computing the one-loop masses via the relevant
Feynman diagrams.  This technique is extremely economical, and
provides results in closed form in full generality, e.g. for general
values of the couplings, etc. The correctness of the original ansatz
for the vacuum can eventually be confirmed by the results of the
computation (namely positive one-loop squared masses, and negligible
tadpoles for the classically massive fields \footnote{Since
  supersymmetry is spontaneously broken the effective potential will
  get renormalized by quantum effects, and thus classically massive
  fields might shift slightly. This appears as a one loop tadpole
  which can be encoded as a small shift of $\mu$. This will enter in
  the two loop computation of the pseudomoduli masses, which are
  beyond the scope of the present paper.}).
  
Hence, our strategy to study the one-loop stabilization in this paper
is as follows:
\begin{itemize}
\item First we choose an ansatz for the classical minimum to become
  the one-loop vacuum.  It is natural to propose a point of maximal
  enhanced symmetry (in particular, close to the origin in the space
  of vevs for $M$ there exist and R-symmetry, whose breaking by gauge interactions (via anomalies) is negligible in that region). Hence the natural candidate for the one-loop minimum is
  \begin{equation}
    q = \tilde q ^{\mathrm{T}} = \left(\begin{array}{c}
        \mu \\ 0
      \end{array}\right),
    \label{minimum}
  \end{equation}
  with the rest of the fields set to 0. This initial ansatz for the
  one-loop minimum is eventually confirmed by the positive square
  masses at one-loop resulting from the computations described below.
  In our more general discussion of meta-stable minima in runaway
  quiver gauge theories, our ansatz for the one-loop minimum is a
  direct generalization of the above (and is similarly eventually
  confirmed by the one-loop mass computation).
\item Then we expand the field linearly around this vacuum, and
  identify the set of classically massless fields. We refer to these
  as pseudomoduli (with some abuse of language, since there could be
  massless fields which are not classically flat directions due to
  higher potential terms)
\item As a final step we compute one-loop masses for these
  pseudomoduli by evaluating their two-point functions via
  conventional Feynman diagrams, as explained in more detail in
  appendix \ref{basicamplitudes} and illustrated below in several
  examples.
\end{itemize}

The ISS model is a simple example where this technique can be
illustrated. Considering the above ansatz for the vacuum, we expand
the fields around this point as:
\begin{equation}
\label{eq:iss-linear-expansion}
q = \left(\begin{array}{c}\mu + \frac{1}{\sqrt{2}}(\xi_+ + \xi_-) \\
\frac{1}{\sqrt{2}}(\rho_+ + \rho_-)
\end{array}\right),\qquad
\tilde q^{\mathrm T} = \left(\begin{array}{c}\mu +
    \frac{1}{\sqrt{2}}(\xi_+ - \xi_-) \\
\frac{1}{\sqrt{2}}(\rho_+ - \rho_-)
\end{array}\right),\qquad M = \left(\begin{array}{cc}
Y & Z \\
\tilde Z^{\mathrm T} & \Phi
\end{array}\right),
\end{equation}
where we have taken linear combinations of the fields in such a way
that the bosonic mass matrix is diagonal. This will also be convenient
in section \ref{sec:goldstones}, where we discuss the Goldstone bosons
in greater detail. 

We now expand the superpotential (\ref{eq:iss-W-magnetic}) to get 
\begin{eqnarray}
\nonumber W & = & \sqrt{2}\mu \xi_+Y + \frac{1}{\sqrt{2}}\mu Z\rho_+ +
\frac{1}{\sqrt{2}}\mu Z\rho_-
+ \frac{1}{\sqrt{2}}\mu \rho_+\tilde Z -  \frac{1}{\sqrt{2}}\mu
\rho_-\tilde Z \\
&& + \frac{1}{2} \rho_+^2 \Phi - \frac{1}{2} \rho_-^2 \Phi - \mu^2\Phi
+ \ldots,
\end{eqnarray}
where we have not displayed terms of order three or higher in the
fluctuations, unless they contain $\Phi$, since they are irrelevant
for the one loop computation we will perform. Note also that we have
set $h=1$ and we have removed the trace (the matricial structure is
easy to restore later on, here we just set $N_f=2$ for simplicity).
The massless bosonic fluctuations are given by $\re\rho_+$,
$\im\rho_-$, $\Phi$ and $\xi_-$. The first two together with $\im
\xi_-$ are Goldstone bosons, as explained in section
\ref{sec:goldstones}. Thus the pseudomoduli we are interested in are given
by $\Phi$ and $\re \xi_-$. Let us focus on $\Phi$ (the case of
$\re\xi_-$ admits a similar discussion). In this case the relevant
terms in the superpotential simplify further, and just the following
superpotential contributes:
\begin{eqnarray}
\nonumber W & = & \mu Z\frac{1}{\sqrt{2}}(\rho_+ + \rho_-)
+ \mu \tilde Z \frac{1}{\sqrt{2}}(\rho_+ - \rho_-)  + \frac{1}{2} \rho_+^2 \Phi - \frac{1}{2} \rho_-^2 \Phi - \mu^2\Phi
+ \ldots,
\end{eqnarray}
which we recognize, up to a field redefinition, as the symmetric model
of appendix \ref{sec:basic-superpotentials}. We can thus directly read the
result
\begin{equation}
\delta m^2_\Phi = \frac{|h|^4\mu^2}{8\pi^2}(\log 4 - 1).
\end{equation}
This matches the value given in \cite{Intriligator:2006dd}, which
was found using the Coleman-Weinberg potential.

\subsection{The Goldstone bosons}
\label{sec:goldstones}

\begin{figure}[!htp]
  \begin{center}
    \input{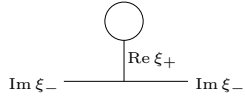}
  \end{center}
  \caption{\small Schematic tadpole contribution to the $\im\xi_-$ two
    point function. Both bosons and fermions run in the loop.}
  \label{fig:diagram}
\end{figure}

One aspect of our technique that merits some additional explanation
concerns the Goldstone bosons. The one-loop computation of the masses
for the fluctuations associated to the symmetries broken by the
vacuum, using just the interactions described in appendix
\ref{basicamplitudes}, leads to a non-vanishing result.  This puzzle
is however easily solved by realizing that certain (classically
massive) fields have a one-loop tadpole. This leads to a new
contribution to the one-loop Goldstone two-point amplitude, given by
the diagram in Figure \ref{fig:diagram}. Adding this contribution the
total one-loop mass for the Goldstone bosons is indeed vanishing, as
expected. This tadpole does not affect the computation of the one-loop
pseudomoduli masses (except for $\re\xi_+$, but its mass remains
positive) as it is straightforward to check.

The structure of this cancellation can be understood by using the
derivation of the Goldstone theorem for the 1PI effective potential, as we now discuss.
The proof can be found in slightly more detail, together with other
proofs, in \cite{Weinberg:1996kr}. Let us denote by $V$ the 1PI
effective potential.  Invariance of the action under a given symmetry
implies that
\begin{equation}
\frac{\delta V}{\delta \phi_i} \Delta \phi_i = 0,
\end{equation}
where we denote by $\Delta \phi_i$ the variation of the field $\phi_i$
under the symmetry, which will in general be a function of all the
fields in the theory. Taking the derivative of this equation with
respect to some other field $\phi_k$
\begin{equation}
\label{eq:GoldstoneTheorem}
\frac{\delta^2 V}{\delta\phi_i \delta\phi_k} \Delta \phi_i +
\frac{\delta V}{\delta\phi_i}\cdot \frac{\delta\Delta\phi_i}{\delta\phi_k}
= 0.
\end{equation}
Let us consider how this applies to our case. At tree level, there is
no tadpole and the above equation (truncated at tree level) states
that for each symmetry generator broken by the vacuum, the value of
$\Delta\phi_i$ gives a nonvanishing eigenvector of the mass matrix
with zero eigenvalue. This is the classical version of the Goldstone
theorem, which allows the identification of the Goldstone bosons of
the theory.

For instance, in the ISS model in the previous section (for $N_f=2$),
there are three global symmetry generators broken at the minimum
described around (\ref{minimum}).  The $SU(2)\times U(1)$ symmetry of
the potential gets broken down to a $U(1)'$, which can be understood
as a combination of the original $U(1)$ and the $t_z$ generator of
$SU(2)$.  The Goldstone bosons can be taken to be the ones associated
to the three generators of $SU(2)$, and correspond (for $\mu$ real) to
$\im\xi_-$, $\im \rho_-$ and $\re \rho_+$, in the parametrization of
the fields given by equation (\ref{eq:iss-linear-expansion}).

Even in the absence of tree-level tadpoles, there could still be a
one-loop tadpole. When this happens, there should also be a
non-trivial contribution to the mass term for the Goldstone bosons in
the one-loop 1PI potential, related to the tadpole by the one-loop
version of (\ref{eq:GoldstoneTheorem}). This relation guarantees that
the mass term in the physical (i.e. Wilsonian) effective potential,
which includes the 1PI contribution, plus those of the diagram in
Figure \ref{fig:diagram}, vanishes, as we described above.

In fact, in the ISS example, there is a non-vanishing one-loop tadpole
for the real part of $\xi_+$ (and no tadpole for other fields). The
calculation of the tadpole at one loop is straightforward, and we will
only present here the result
\begin{equation}
i\cam = \frac{-i |h|^4\mu^3}{(4\pi)^2}(2\log 2).
\end{equation}
The 1PI one-loop contribution to the Goldstone boson mass is also simple to
calculate, giving the result
\begin{equation}
i\cam = \frac{-i |h|^4\mu^2}{(4\pi)^2}(\log 2).
\end{equation}
Using the variations of the relevant fields under the symmetry
generator, e.g. for  $t_z$,
\begin{eqnarray}
\Delta\re \xi_+ & = & - \im \xi_- \\
\Delta\im \xi_- & = & \re \xi_+ + 2\mu.
\end{eqnarray}
we find that the (\ref{eq:GoldstoneTheorem}) is satisfied at one-loop.
\begin{equation}
\left\langle \frac{\delta^2 V}{\delta\phi_i \delta\phi_k} \Delta \phi_i +
\frac{\delta V}{\delta\phi_i}\cdot
\frac{\delta\Delta\phi_i}{\delta\phi_k}\right\rangle =
m^2_{\im \xi_-}\cdot 2\mu + (\re
\xi_+ \mathrm{ tadpole})\cdot (-1) = 0.
\end{equation}
A very similar discussion applies to $t_x$ and $t_y$.

\smallskip

The above discussion of Goldstone bosons can be similarly carried out
in all examples of this paper. Hence, it will be enough to carry out
the computation of the 1PI diagrams discussed in appendix
\ref{basicamplitudes}, and verify that they lead to positive squared
masses for all classically massless fields (with Goldstone bosons
rendered massless by the additional diagrams involving the tadpole).

\section{Meta-stable vacua in quiver gauge theories with DSB branes}
\label{dsb}

In this section we show the existence of a meta-stable vacuum in a few
examples of gauge theories on DSB branes, upon the addition of massive
flavors. As already discussed in \cite{Franco:2006es}, the choice of fractional branes of DSB kind is crucial in the result. The reason is that in order to
have the ISS structure, and in particular supersymmetry breaking by
the rank condition, one needs a node such that its Seiberg dual
satisfies $N_{f}>N$, with $N=N_f-N_c$ with $N_c$, $N_f$ the number of
colors, flavors of that gauge factor. Denoting $N_{f,0}$, $N_{f,1}$
the number of massless and massive flavors (namely flavors arising
from bi-fundamentals of the original D3-brane quiver, or introduced by
the D7-branes), the condition is equivalent to $N_{f,0}<N_c$. This is
precisely the condition that an ADS superpotential is generated, and
is the prototypical behavior of DSB branes
\cite{Berenstein:2005xa,Franco:2005zu,Bertolini:2005di,Brini:2006ej}.

Another important general comment, also discussed in
\cite{Franco:2006es}, is that theories on DSB branes generically
contain one or more chiral multiplets which do not appear in the
superpotential. Being decoupled, such fields remain as accidental flat
directions at one-loop, so that the one-loop minimum is not isolated.
The proper treatment of these flat directions is beyond the reach of
present tools, so they remain an open question. However, it is
plausible that they do not induce a runaway behavior to infinity,
since they parametrize a direction orthogonal to the fields
parametrizing the runaway of DSB fractional branes.

\subsection{The complex cone over $dP_1$}

In this section we describe the most familiar example of quiver gauge
theory with DSB fractional branes, the $dP_1$ theory. In this theory,
a non-perturbative superpotential removes the classical supersymmetric
vacua
\cite{Berenstein:2005xa,Franco:2005zu,Bertolini:2005di}.
Assuming canonical K\"ahler potential the theory has a runaway behavior
\cite{Franco:2005zu,Intriligator:2005aw}. In this section, we
revisit with our techniques the result in \cite{Franco:2006es} that
the addition of massive flavors can induce the appearance of
meta-stable supersymmetry breaking minima, long-lived against
tunneling to the runaway regime.  As we show in coming sections, this
behavior is prototypical and extends to many other theories with DSB
fractional branes. The example is also representative of the
computations for a general quiver coming from a brane at a toric
singularity, and illustrates the usefulness of the direct Feynman
diagram evaluation of one-loop masses.

\begin{figure}[!htp]
\centering
\psfrag{3i}{$Q_{3i}$}
\psfrag{i2}{${\tilde Q}_{i2}$}
\psfrag{2j}{$Q_{2j}$}
\psfrag{j1}{${\tilde Q}_{j1}$}
\psfrag{1k}{$Q_{1k}$}
\psfrag{k3}{${\tilde Q}_{k3}$}
\includegraphics[scale=0.70]{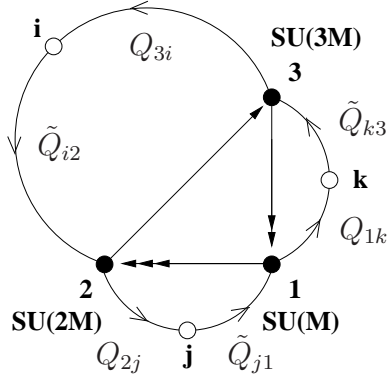}
\caption{\small Extended quiver diagram for a $dP_1$ theory with
  flavors, from \cite{Franco:2006es}.}
\label{fig:extd3d7}
\end{figure}

Consider the $dP_1$ theory, realized on a set of $M$ fractional
D3-branes at the complex cone over $dP_1$. In order to introduce
additional flavors, we introduce sets of $N_{f,1}$ D7-branes wrapping
non-compact 4-cycles on the geometry and passing through the singular
point. We refer the reader to \cite{Franco:2006es}, and also to later
sections, for more details on the construction of the theory, and in
particular on the introduction of the D7-branes. Its quiver is shown
in Figure \ref{fig:extd3d7}, and its superpotential is
\begin{eqnarray}
  W & = & \lambda (X_{23}X_{31}Y_{12} - X_{23} Y_{31} X_{12}) \nonumber \\
  &+& \lambda' (Q_{3i} {\tilde Q}_{i2} X_{23} + Q_{2j} {\tilde Q}_{j1}
  X_{12} + Q_{1k} {\tilde Q}_{k3} X_{31}) \nonumber \\
  &+&m_3 Q_{3i} {\tilde Q}_{k3} \delta_{ik} +
  m_2 Q_{2j} {\tilde Q}_{i2} \delta_{ji} +
  m_1 Q_{1k} {\tilde Q}_{j1} \delta_{kj},
\end{eqnarray} 
where the subindices denote the groups under which the field is
charged. The first line is the superpotential of the theory of
fractional brane, the second line describes 77-73-37 couplings between
the flavor branes and the fractional brane, and the last line gives
the flavor masses.  Note that there is a massless field, denoted
$Z_{12}$ in \cite{Franco:2006es}, that does not appear in the
superpotential. This is one of the decoupled fields mentioned above,
and we leave its treatment as an open question.

We are interested in gauge factors in the free magnetic phase. This is
the case for the $SU(3M)$ gauge factor in the regime
\begin{equation}
M+1 \leq N_{f,1} < \frac{5}{2}M.
\end{equation} 
To apply Seiberg duality on node 3, we introduce
the dual mesons:
\begin{equation}
\begin{array}{rlcrl}
M_{21} & =\, {1\over \Lambda} X_{23} X_{31} & \quad ; \quad  & N_{k1} & 
= \, {1\over \Lambda} {\tilde Q}_{k3}X_{31} \\
M_{21}' & =\, {1\over \Lambda} X_{23} Y_{31} & \quad ; \quad  & N_{k1}' & = \, 
{1 \over \Lambda} {\tilde Q}_{k3}Y_{31} \\
N_{2i} & =\, {1\over \Lambda} X_{23} Q_{3i} & \quad ; \quad  & \Phi_{ki} & = 
\, {1\over \Lambda} {\tilde Q}_{k3}Q_{3i}
\end{array}
\end{equation}
and we also replace the electric quarks $Q_{3i}$, ${\tilde Q}_{k3}$,
$X_{23}$, $X_{31}$, $Y_{31}$ by their magnetic duals ${\tilde
  Q}_{i3}$, $Q_{3k}$, $X_{32}$, $X_{13}$, $Y_{13}$. The magnetic
superpotential is given by rewriting the confined fields in terms of
the mesons and adding the coupling between the mesons and the dual
quarks,
\begin{eqnarray}
W & = & h\, (\, M_{21} X_{13} X_{32} \, + \, M_{21}' Y_{13} X_{32} 
\, +\,
N_{2i} {\tilde Q}_{i3} X_{32} \nonumber \\
& + & N_{k1}X_{13}Q_{3k} \, + \, N_{k1}' Y_{13} Q_{3k} \, +\, \Phi_{ki}
{\tilde Q}_{i3} Q_{3k}\, ) \nonumber \\
& + &
h\mu_0\,(\, M_{21} Y_{12} \, -\, M_{21}' X_{12} \,)\, +\, 
\mu'\, Q_{1k} N_{k1}\, +\, \mu'\, N_{2i} {\tilde Q}_{i2}\nonumber \\
& - & h\mu^{\, 2} \Tr\Phi \, 
+\, \lambda' \, Q_{2j} {\tilde Q}_{j1} X_{12} \, 
+\, m_2 Q_{2i}{\tilde Q}_{i2} \, +\, m_1 Q_{1i}{\tilde Q}_{i1}.
\end{eqnarray}
This is the theory we want to study. In order to simplify the
treatment of this example we will disregard any subleading terms
in $m_i/\mu'$, and effectively integrate out $N_{k1}$ and
$N_{2i}$ by substituting them by 0. This is not necessary, and
indeed the computations in the next sections are exact. We do it here in order to  compare results with \cite{Franco:2006es}.

As in the ISS model, this theory breaks supersymmetry via the rank
condition. The fields $\tilde Q_{i3}$, $Q_{3k}$ and
$\Phi_{ki}$ are the analogs of $q$, $\tilde q$ and $M$ in the ISS case
discussed above. This motivates a vacuum ansatz analogous to (\ref{minimum}) and the following linear expansion:
\begin{equation}
  \begin{array}{ccccc}
    \Phi = \left(\begin{array}{cc}
        \phi_{00} & \phi_{01} \\
        \phi_{10} & \phi_{11}
      \end{array}\right) & ; &
    \tilde Q_{i3} = \left(\begin{array}{c}
        \mu e^{\theta} + Q_{3,1} \\
        \tilde Q_{3,2}
      \end{array}\right) & ; &
    Q_{3i}^{\mathrm{T}} = \left(\begin{array}{c}
        \mu e^{-\theta} + Q_{3,1} \\
        Q_{3,2}
      \end{array}\right) \\
    \tilde Q_{k1} = \left(\begin{array}{c}
        \tilde Q_{1,1} \\ y
      \end{array}\right) & ; &
    Q_{2j} = \left(\begin{array}{cc}
        Q_{2, 11} & x \\
        Q_{2, 21} & x'
      \end{array}\right) & ; &
    M_{21} = \left(\begin{array}{c}
        M_{21,1} \\ M_{21,2}
      \end{array}\right) \\
    Y_{13}=\left(Y_{13}\right) & ; &
    X_{12}^\mathrm{T} = \left(\begin{array}{c}
        X_{12,1} \\ X_{12,2}
      \end{array}\right) & ; &
    X_{32}^\mathrm{T} = \left(\begin{array}{c}
        X_{32,1} \\ X_{32, 2}
      \end{array}\right) \\
    Y_{12}^\mathrm{T} = \left(\begin{array}{c}
        Y_{12,1} \\ Y_{12,2}
        \end{array}\right) & ; &
    N'_{k1} = \left(\begin{array}{c}
        N'_{k1,1} \\ z
      \end{array}\right) & ; &
    M'_{21} = \frac{\lambda'}{h\mu_0}\left(\begin{array}{c}
        M'_{21,1} \\ M'_{21,2}
      \end{array}\right) \\
    &&
    X_{13} = \left(X_{13}\right).
  \end{array}
\end{equation}
Note that we have chosen to introduce the nonlinear expansion in
$\theta$ in order to reproduce the results found in the literature in
their exact form\footnote{A linear expansion would lead to identical
  conclusions concerning the existence of the meta-stable vacua, but
  to one-loop masses not directly amenable to comparison with results
  in the literature.}. Note also that for the sake of clarity we have not been explicit about
the ranks of the different matrices. They can be easily worked out (or for this case, looked up in \cite{Franco:2006es}), and we will restrict ourselves to the 2
flavor case where the matrix structure is trivial. As a last
remark, we are not being explicit either about the definitions of the
different couplings in terms of the electric theory. This can be done
easily (and as in the ISS case they involve an unknown coefficient in
the K\"ahler potential), but in any event, the existence of the
meta-stable vacua can be established for general values of the
coefficients in the superpotential. Hence we skip this more detailed
but not very relevant discussion.

The next step consists in expanding the superpotential and identifying
the massless fields. We get the following quadratic contributions to
the superpotential:
\begin{eqnarray}
  W_{\mathrm{mass}} & = & 2 h\mu\phi_{00}\tilde Q_{3,1} + h\mu
  \phi_{01}\tilde Q_{3,2} + h\mu\phi_{10}Q_{3,2} \nonumber \\
  & + & h\mu_0 M_{21,1} Y_{12,1} + h\mu_0 M_{21,2} Y_{12,2}
  - \lambda' M'_{21,1} X_{12,1} - \lambda' M'_{21,2} X_{12,2} \nonumber \\
  & + & h\mu N'_{k1,1} Y_{13} - h_1\mu \tilde Q_{1,1} X_{13} -
  h_2\mu Q_{2,11} X_{32,1} - h_2\mu Q_{2,21} X_{32,2}.
\end{eqnarray}
The fields massless at tree level are $x$, $x'$, $y$, $z$, $\phi_{11}$, $\theta$, $Q_{3,2}$ and $\tilde Q_{3,2}$. Three of these are Goldstone bosons as described in
the previous section. For real $\mu$ they are $\im\theta$,
$\re(\tilde Q_{3,2} + Q_{3,2})$ and $\im(\tilde Q_{3,2} - Q_{3,2})$.
We now show that all other classically massless fields get masses at one loop (with positive squared masses). 

As a first step towards finding the one-loop correction, notice that
the supersymmetry breaking mechanism is extremely similar to the one
in the ISS model before, in particular it comes only from the
following couplings in the superpotential:
\begin{equation}
  W_{\mathrm{rank}} = h Q_{3,2} \tilde Q_{3,2} \phi_{11} - h\mu^2
  \phi_{11} + \ldots
\end{equation}
This breaks the spectrum degeneracy in the multiplets $Q_{3,2}$ and $\tilde
Q_{3,2}$ at tree level, so we refer to them as the fields with broken supersymmetry. 

Let us  compute now the correction for the mass of $x$, for
example. For the one-loop computation we just need the cubic terms involving one pseudomodulus and at least one of the broken supersymmetry fields, and any quadratic term involving fields present in the previous set of couplings. From the complete expansion one finds the following supersymmetry breaking sector:
\begin{equation}
  W_{\mathrm{symm.}} = h \phi_{11} Q_{3,2} \tilde Q_{3,2} +
  h\mu\phi_{01}\tilde Q_{3,2} + h\mu \phi_{10} Q_{3,2} - h\mu^2
  \phi_{11}.
\end{equation}
The only cubic term involving the pseudomodulus $x$ and the broken supersymmetry fields is
\begin{equation}
  W_{\mathrm{cubic}} = -h_2\, x\, \tilde Q_{3,2} X_{32,1},
\end{equation}
and there is a quadratic term involving the field  $X_{32,1}$ 
\begin{equation}
  W_{\mathrm{mass\,coupling}} = -h_2\mu Q_{2,11} X_{32,1}.
\end{equation}
Assembling the three previous equations, the resulting superpotential corresponds to the {\em
  asymmetric} model in appendix \ref{sec:basic-superpotentials}, so we can directly obtain the one-loop mass for $x$:
\begin{equation}
\delta m^2_x = \frac{1}{16\pi^2}|h|^4\mu^2
\cc\left(\frac{|h_2|^2}{|h|^2}\right).
\end{equation}
Proceeding in a similar way, the one-loop masses for $\phi_{11}$,
$x'$, $y$ and $z$ are:
\begin{eqnarray}
\delta m^2_{\phi_{11}} & = & \frac{1}{8\pi^2}|h|^4\mu^2(\log 4 - 1)
\nonumber \\
\delta m^2_{x'} & = & \frac{1}{16\pi^2}|h|^4\mu^2
\cc\left(\frac{|h_2|^2}{|h|^2}\right), \nonumber \\
\delta m^2_y & = &
\frac{1}{16\pi^2}|h|^4\mu^2\cc\left(\frac{|h_1|^2}{|h|^2}\right)
\nonumber \\
\delta m^2_z & = & \frac{1}{16\pi^2}|h|^4\mu^2(\log 4 - 1).
\end{eqnarray}

There is just one pseudomodulus left, $\re\theta$, which is
qualitatively different to the others. With similar reasoning, one
concludes that it is necessary to study a superpotential of the form
\begin{equation}
  W = h(X\phi_1\phi_2 + \mu e^{\theta}\phi_1 \phi_3
  +\mu e^{-\theta}\phi_2\phi_4 - \mu^2 X).
\end{equation}
Due to the non-linear parametrization, the expansion in $\theta$ shows that there is a term quadratic in $\theta$ which contributes to the one-loop mass via a vertex with two bosons and two fermions, the relevant diagram is shown in Figure
\ref{feynman_diagrams}d. The result is a vanishing mass for ${\im
  \theta}$, as expected for a Goldstone boson (the one-loop tadpole
vanishes in this case), and a non-vanishing mass for ${\re\theta}$
\begin{equation}
\delta m^2_{\re\theta} = \frac{1}{4\pi^2}|h|^4 \mu^4(\log 4 - 1).
\end{equation}

We conclude by mentioning that all squared masses are positive, thus
confirming that the proposed point in field space is the one-loop
minimum. As shown in \cite{Franco:2006es}, this minimum is
parametrically long-lived against tunneling to the runaway regime.

\subsection{Additional examples: The $dP_2$ case}

\begin{figure}[!htp]
  \epsfxsize = 4cm
  \centerline{\epsfbox{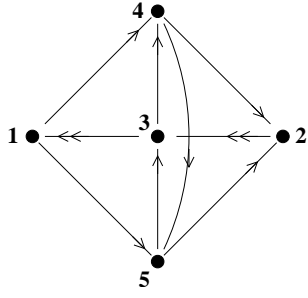}}
  \caption{\small Quiver diagram for the $dP_2$ theory.}
  \label{quiverdp2}
\end{figure}

\begin{figure}[!htp]
  \epsfxsize = 4cm
  \centerline{\epsfbox{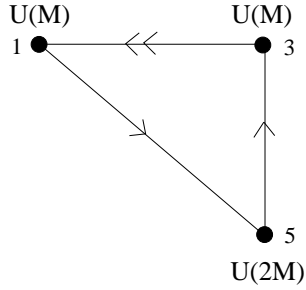}}
  \caption{\small Quiver diagram for the $dP_2$ theory with $M$
 DSB  fractional branes.}
  \label{dp2_frac}
\end{figure}

Let us apply these techniques to consider new examples. In this
section we consider a DSB fractional brane in the complex cone over
$dP_2$, which provides another quiver theory with runaway behavior
\cite{Franco:2005zu}.  The quiver diagram for $dP_2$ is given in
Figure \ref{quiverdp2}, with superpotential
\begin{eqnarray} W & = &
X_{34}X_{45}X_{53}-X_{53}Y_{31}X_{15}-X_{34}X_{42}Y_{23} +
Y_{23}X_{31}X_{15}X_{52}
\nonumber\\
& + & X_{42}X_{23}Y_{31}X_{14}-X_{23}X_{31}X_{14}X_{45}X_{52}
\label{W_dP2_1}
\end{eqnarray} 
We consider a set of $M$ DSB fractional branes, corresponding to choosing ranks
$(M,0,M,0,2M)$ for the corresponding gauge factors. The resulting quiver is shown in Figure \ref{dp2_frac}, with superpotential 
\begin{eqnarray}
W = -\lambda X_{53}Y_{31}X_{15}
\label{W_dP2_2}
\end{eqnarray}%

\begin{figure}[!htp]
    \centering
    \psfrag{Q1i}{$Q_{1i}$}
    \psfrag{Qi3}{$Q_{i3}$}
    \psfrag{Q3j}{$Q_{3j}$}
    \psfrag{Qj5}{$Q_{j5}$}
    \psfrag{Q5k}{$Q_{5k}$}
    \psfrag{Qk1}{$Q_{k1}$}
    \includegraphics[scale=3.00]{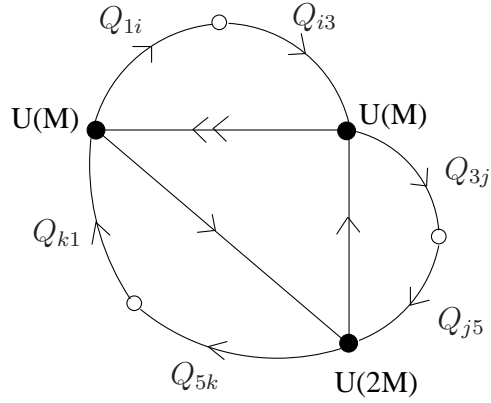}
  \caption{\small Quiver for the $dP_2$ theory with $M$ fractional branes and flavors.}
  \label{dp2_flav}
\end{figure}

Following \cite{Franco:2006es} and appendix \ref{Riemannsurface}, one can introduce D7-branes leading to D3-D7 open strings providing (possibly massive) flavors for all gauge factors, and having cubic couplings with diverse D3-D3 bifundamental chiral multiplets. We obtain the quiver in Figure \ref{dp2_flav}.
Adding the cubic 33-37-73 coupling superpotential, and the flavor masses, the complete superpotential reads
\begin{eqnarray} W_{total} & = & -\lambda
X_{53}Y_{31}X_{15}-\lambda'(Q_{1i}\tilde{Q}_{i3}Y_{31}+Q_{3j}\tilde{Q}_{j5}X_{53}+
Q_{5k}\tilde{Q}_{k1}X_{15})
\nonumber\\
& + & m_{1}Q_{1i}\tilde{Q}_{k1} + m_{2}Q_{3j}\tilde{Q}_{i3} +
m_{5}Q_{5k}\tilde{Q}_{j5}
\label{W_dP2_total}
\end{eqnarray}
where $1,2,3$ are the gauge group indices and $i,j,k$ are the flavor indices.

We consider the $U(2M)$ node in the free magnetic phase, namely
\begin{eqnarray}
M+1 & \leq & N_{f,1} \, < \, 2M 
\end{eqnarray} 
After Seiberg Duality the dual gauge factor is $SU(N)$ with $N=N_{f,1}-M$ and dynamical scale $\Lambda$. To get
the matter content in the dual, we replace the microscopic flavors
$Q_{5k}$, ${\tilde Q}_{j5}$, $X_{53}$, $X_{15}$ by the dual flavors
${\tilde Q}_{k5}$, $Q_{5j}$, $X_{35}$, $X_{51}$ respectively. We also
have the mesons related to the fields in the electric theory by \begin{equation}
\begin{array}{rlcrl}
  M_{1k} & =\, {1\over \Lambda} X_{15} Q_{5K} & \quad ; \quad  & {\tilde N}_{j3} & 
  = \, {1\over \Lambda} {\tilde Q}_{j5}X_{53} \\
  M_{13} & =\, {1\over \Lambda} X_{15} X_{53} & \quad ; \quad  & {\tilde \Phi}_{jk} & = \, 
  {1 \over \Lambda} {\tilde Q}_{j5}Q_{5k} 
\end{array}
\end{equation}
There is a cubic superpotential coupling the mesons and the dual flavors
\begin{eqnarray}
W_{mes.} & = & h\, (\, M_{1k} {\tilde Q}_{k5} X_{51} \, + \, M_{13} X_{35} X_{51} 
\, +\,
{\tilde N}_{j3} X_{35} Q_{5j}  \, + \, {\tilde \Phi}_{jk} {\tilde Q}_{k5 }Q_{5j} \, )
\end{eqnarray}
where $h\, =\, \Lambda/\hat\Lambda$ with $\hat\Lambda$ given by
$\Lambda_{elect}^{\, 3N_c-N_f}\, \Lambda^{3(N_f-N_c)-N_f}\, =\, 
\hat \Lambda^{N_f}$, where $\Lambda_{elect}$ is the dynamical scale of the electric theory.
Writing the classical superpotential terms of the new fields gives
\begin{eqnarray}
W_{clas.} & = & - \, h \, \mu_0\, M_{13} Y_{31} \, +\, 
 \lambda' \, Q_{1i} {\tilde Q}_{i3} Y_{31} \,
 +\, \mu'\, {\tilde N}_{j3} Q_{3j} \, + \, \mu'\, M_{1k} {\tilde Q}_{k1} 
\, 
\nonumber \\
& + & \, m_1 Q_{1i}{\tilde Q}_{k1} \, 
+\, m_3 Q_{3j}{\tilde Q}_{i3} \, - \, h\mu^{\, 2} \Tr\Phi
\end{eqnarray}
where $\mu_0=\lambda \Lambda$, $\mu'=\lambda'\Lambda$, and
$\mu^{\,2}=-m_5\hat\Lambda$.
So the complete superpotential in the Seiberg dual is
\begin{eqnarray}
W_{dual} & = & - \, h \, \mu_0\, M_{13} Y_{31} \, +\, 
 \lambda' \, Q_{1i} {\tilde Q}_{i3} Y_{31} \,
 +\, \mu'\, {\tilde N}_{j3} Q_{3j} \, + \, \mu'\, M_{1k} {\tilde Q}_{k1} 
\, 
\nonumber \\
& + & \, m_1 Q_{1i}{\tilde Q}_{k1} \, 
+\, m_3 Q_{3j}{\tilde Q}_{i3} \, - \, h\mu^{\, 2} \Tr\Phi
\nonumber \\
& + & \, h\, (\, M_{1k} {\tilde Q}_{k5} X_{51} \, + \, M_{13} X_{35} X_{51} 
\, +\,
{\tilde N}_{j3} X_{35} Q_{5j}  \, + \, {\tilde \Phi}_{jk} {\tilde Q}_{k5 }Q_{5j} \, )
\label{Wtot}
\end{eqnarray} 

This superpotential has a sector completely analogous to the ISS model, triggering supersymmetry breaking by the rank condition. This suggests the following ansatz for the point to become the one-loop vacuum
 \begin{equation}
Q_{5k}  = \tilde Q_{5k}^{\, \mathrm{T}} = \left(\begin{array}{c}
        \mu \\ 0
      \end{array}\right),
     \end{equation}
with all other vevs set to zero. Following our technique as explained above, we expand fields at linear order around this point. Focusing on $N_{f,1} = 2$ and $N_c =1$ for simplicity (the general case can be easily recovered), we have
\begin{equation}
\begin{array}{c}
\begin{array}{ccccc}
{\tilde Q}_{k5}\, =\, \pmatrix{\mu + \delta {\tilde Q}_{5,1} \cr \delta {\tilde Q}_{5,2}} & \ \ ; \ \ & Q_{5k}\, =\, ( \mu + \delta Q_{5,1} \, ; \, \delta Q_{5,2}) & \ \ ; \ \ & \Phi\, =\, \pmatrix{\delta \Phi_{0,0} & \delta \Phi_{0,1} \cr \delta \Phi_{1,0} & \delta \Phi_{1,1}} 
\end{array} \\ \\
\begin{array}{ccccccc}
{\tilde Q}_{k1}\, =\, \pmatrix{ \delta {\tilde Q}_{1,1} \cr \delta {\tilde Q}_{1,2}}  & ; \ 
Q_{1i}\, =\, ( \delta Q_{1,1} \, ; \, \delta Q_{1,2})  & ; \   {\tilde Q}_{i3}\, =\, \pmatrix{ \delta {\tilde Q}_{3,1} \cr \delta {\tilde Q}_{3,2}}  & ; \ 
Q_{3j}\, =\, ( \delta Q_{3,1} \, ; \, \delta Q_{3,2}) 
\end{array} \\ \\
\begin{array}{cccccc}
{\tilde N}_{j3}\, =\, \pmatrix{ \delta {\tilde N}_{3,1} \cr \delta {\tilde N}_{3,2}} & ; \ 
M_{1k}\, =\, ( \delta M_{1,1} \, ; \, \delta M_{1,2}) & ; \ M_{13} = \delta M_{13} &  ; \ Y_{31} = \delta Y_{31} & ; \  X_{51} = \delta X_{51} 
\end{array} \\ \\ 
 X_{35} = \delta X_{35}    
\end{array}
\end{equation}
Inserting this into equation (\ref{Wtot}) gives
\begin{eqnarray}
W_{dual} & = & - \, h \, \mu_0\, \delta M_{13} \delta Y_{31} \, +\, 
 \lambda' \, \delta Q_{1,1} \delta {\tilde Q}_{3,1}  \delta Y_{31} \,
 +\, \lambda' \, \delta Q_{1,2} \delta {\tilde Q}_{3,2}  \delta Y_{31} \, 
\nonumber \\
& + &\, \mu'\, \delta {\tilde N}_{3,1} \delta Q_{3,1} \, +\, \mu'\, \delta {\tilde N}_{3,2}     \delta Q_{3,2} \, 
 + \, \mu'\, \delta M_{1,1} \delta {\tilde Q}_{1,1} \, + \, \mu'\, \delta M_{1,2} \delta {\tilde Q}_{1,2} \,
\nonumber \\
& + & \, m_1 \delta Q_{1,1} \delta {\tilde Q}_{1,1} \, +  \, m_1 \delta Q_{1,2} \delta {\tilde Q}_{1,2} \,
+\, m_3 \delta Q_{3,1} \delta {\tilde Q}_{3,1} \, +\, m_3 \delta Q_{3,2} \delta {\tilde Q}_{3,2} \, 
\nonumber \\
& - & \, h\mu^{\, 2} \delta \Phi_{11}
 +  \, h\, (\, \mu \delta M_{1,1} \delta X_{51} \, + \, \delta M_{1,1} \delta {\tilde Q}_{5,1} \delta X_{51} \, + \, \delta M_{1,2} \delta {\tilde Q}_{5,2} \delta X_{51}   \, 
\nonumber \\
& + & \, \delta M_{13} \delta X_{35} \delta X_{51} 
 +  \, \mu \delta X_{35} \delta {\tilde N}_{3,1} \, + \,
\delta X_{35} \delta {\tilde N}_{3,1}  \delta Q_{5,1}  \, +   \,
\delta X_{35} \delta {\tilde N}_{3,2}  \delta Q_{5,2}  \, 
\nonumber \\
& + & \mu \delta {\tilde Q}_{5,1} \delta \Phi_{00}\, +
\, \mu \delta Q_{5,1} \delta \Phi_{00}   \, +
\, \delta Q_{5,1} \delta {\tilde Q}_{5,1} \delta \Phi_{00} \, +
\, \mu \delta \Phi_{01} \delta {\tilde Q}_{5,2}  \, 
\nonumber \\
& + & \, \delta Q_{5,1} \delta \Phi_{01}\delta {\tilde Q}_{5,2} \, +
\, \mu \delta \Phi_{10}\delta Q_{5,2} 
\, +
\, \delta {\tilde Q}_{5,1} \delta \Phi_{10}\delta Q_{5,2} \, +
\, \delta {\tilde Q}_{5,2} \delta \Phi_{11}\delta Q_{5,2}).
\nonumber
\label{Wtot1}
\end{eqnarray}
We now need to identify the pseudomoduli, in other words the massless
fluctuations at tree level. We focus then just on the quadratic
terms in the superpotential
\begin{eqnarray}
  W_{mass} & = &- \, h \, \mu_0\, \delta M_{13} \delta Y_{31} \nonumber\\
  & + & \, \mu'\, \delta {\tilde N}_{3,1} \delta Q_{3,1}\,+
  m_3 \delta Q_{3,1} \delta {\tilde Q}_{3,1}\,+
  h \mu \delta X_{35} \delta {\tilde N}_{3,1} \nonumber\\
  & + & \, \mu'\, \delta {\tilde N}_{3,2} \delta Q_{3,2} \,+
  m_3 \delta Q_{3,2} \delta {\tilde Q}_{3,2} \nonumber\\
  & + & \, \mu'\, \delta M_{1,1} \delta {\tilde Q}_{1,1} \, +
  \, m_1 \delta Q_{1,1} \delta {\tilde Q}_{1,1} \, +
  h\mu \delta M_{1,1} \delta X_{51} \nonumber\\
  & + & \, \mu'\, \delta M_{1,2} \delta {\tilde Q}_{1,2} \, +
  \, m_1 \delta Q_{1,2} \delta {\tilde Q}_{1,2} \nonumber \\
  & + & \, h\mu \delta {\tilde Q}_{5,1} \delta \Phi_{00}\, +
  \, h\mu \delta Q_{5,1} \delta \Phi_{00} \nonumber\\
  & + & \, h\mu \delta \Phi_{01} \delta {\tilde Q}_{5,2} +
  \, \mu \delta \Phi_{10}\delta Q_{5,2}.
\end{eqnarray}
We have displayed the superpotential so that fields mixing at the quadratic level appear in the same line. In order to identify
the pseudomoduli we have to diagonalize\footnote{As a technical
  remark, let us note that it is possible to set all the mass terms to
  be real by an appropriate redefinition of the fields, so we are
  diagonalizing a real symmetric matrix.} these fields.
Note that the structure of the mass terms corresponds to the one in
appendix \ref{PROOF}, in particular around equation
(\ref{eq:general-mass-mixing}). From the analysis performed there we
know that upon diagonalization, fields mixing in
groups of four (i.e., three mixing terms in the superpotential, for
example the $\delta M_{1,1}$, $\delta {\tilde Q}_{1,1}$, $\delta
Q_{1,1}$, $\delta X_{51}$ mixing) get nonzero masses, while fields mixing in groups of three (two mixing terms in the superpotential, for example $\delta M_{1,2}$, $\delta {\tilde Q}_{1,2}$ and $\delta Q_{1,2}$) give rise to two massive
perturbations and a massless one, a pseudomodulus. We then just need to
study the fate of the pseudomoduli. From the analysis in appendix
\ref{PROOF}, the pseudomoduli coming from the mixing
terms are
\begin{eqnarray}
  Y_1 & = & m_3 \delta {\tilde N}_{3,2} - \mu' \delta {\tilde
    Q}_{3,2}\, ,\nonumber\\
  Y_2 & = & m_1 \delta M_{1,2} - \mu' \delta Q_{1,2}\, , \nonumber\\
  Y_3 & = & h\mu(\delta Q_{5,1} - \delta {\tilde Q}_{5,1})\,.
\end{eqnarray}
In order to continue the analysis, one just needs to change basis to
the diagonal fields and notice that the one loop contributions to the
pseudomoduli are described again by the asymmetric model of appendix
\ref{sec:basic-superpotentials}, so they receive positive definite
contributions. The exact analytic expressions can be easily found with
the help of some computer algebra program, but we omit them here since
they are quite unwieldy.

\section{The general case}
\label{generalcase}

In the previous section we showed that several examples of quiver
gauge theories on DSB fractional branes have metastable vacua once
additional flavors are included. In this section we generalize the
arguments for general DSB branes. We will show how to add D7--branes
in a specific manner so as to generate the appropriate cubic flavor
couplings and mass terms. Once this is achieved, we describe the
structure of the Seiberg dual theory. The results of our analysis show
that, with the specified configuration of D7--branes, the
determination of metastability is greatly simplified and only involves
looking at the original superpotential. Thus, although we do not prove
that DSB branes on arbitrary singularities generate metastable vacua,
we show how one can determine the existence of metastability in a very
simple and systematic manner. Using this analysis we show further
examples of metastable vacua on systems of DSB branes.

\subsection{The general argument}

\subsubsection{Construction of the flavored theories}

\begin{figure}[!htp]
  \centering
    \psfrag{X21}{$X_{21}$}
    \psfrag{Y21}{$Y_{21}$}
    \psfrag{X32}{$X_{32}$}
    \psfrag{Y32}{$Y_{32}$}
    \psfrag{Z32}{$Z_{32}$}
    \psfrag{X14}{$X_{14}$}
    \psfrag{X43}{$X_{43}$}
    \psfrag{Y43}{$Y_{43}$}
  \includegraphics[scale=0.90]{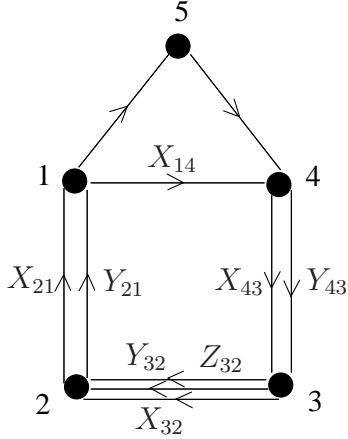}
  \caption{\small Quiver diagram used to illustrate general results.
    It does not correspond to any geometry in particular.}
  \label{arbquiv}
\end{figure}

Consider a general quiver gauge theory arising from branes at
singularities.  As we have argued previously, we focus on DSB branes,
so that there is a gauge factor satisfying $N_{f,0}<N_c$, which can
lead to supersymmetry breaking by the rank condition in its Seiberg
dual. To make the general analysis more concrete, let us consider a
quiver like that in Figure \ref{arbquiv}, which is characteristic
enough, and let us assume that the gauge factor to be dualized
corresponds to node 2. In what follows we analyze the structure of the
fields and couplings in the Seiberg dual, and reduce the problem of
studying the meta-stability of the theory with flavors to analyzing
the structure of the theory in the absence of flavors.

The first step is the introduction of flavors in the theory. As
discussed in \cite{Franco:2006es}, for any bi-fundamental $X_{ab}$ of
the D3-brane quiver gauge theory there exist a supersymmetric
D7-brane leading to flavors $Q_{bi}$, ${\tilde Q}_{ia}$ in the
fundamental (antifundamental) of the $b^{th}$ ($a^{th}$) gauge factor.
There is also a cubic coupling $X_{ab}Q_{bi}{\tilde Q}_{ia}$. Let us
now specify a concrete set of D7-branes to introduce flavors in our
quiver gauge theory. Consider a superpotential coupling of the
D3-brane quiver gauge theory, involving fields charged under the node
to be dualized. This corresponds to a loop in the quiver, involving
node 2, for instance $X_{32}X_{21}X_{14}Y_{43}$ in Figure
\ref{arbquiv}. For any bi-fundamental chiral multiplet in this
coupling, we introduce a set of $N_{f,1}$ of the corresponding
D7-brane. This leads to a set of flavors for the different gauge
factors, in a way consistent with anomaly cancellation, such as that
shown in Figure \ref{arbquiv1}.  The description of this system of
D7-branes in terms of dimer diagrams is carried out in Appendix
\ref{Riemannsurface}. The cubic couplings described above lead to the
superpotential terms\footnote{Here we assume the same coupling, but
  the conclusions hold for arbitrary non-zero couplings.}
\begin{equation} 
W_{flavor} = \lambda' \,( \,
 X_{32}Q_{2b}Q_{b3} \, + \, X_{21}Q_{1a}Q_{a2} \, + \,
X_{14}Q_{4d}Q_{d1}\, + \, Y_{43}Q_{3c}Q_{c4}  \,) 
\end{equation}
Finally, we introduce mass terms for all flavors of all involved gauge
factors:
\begin{equation}
W_{mass} = m_2 Q_{a2}Q_{2b} \, + \, m_3 Q_{b3}Q_{3c} \, + \, m_4
Q_{c4}Q_{4d}\, + \, m_1 Q_{d1}Q_{1a}
\end{equation}
These mass terms break the flavor group into a diagonal subgroup.

\begin{figure}[!htp]
    \centering
    \psfrag{X21}{$X_{21}$}
    \psfrag{Y21}{$Y_{21}$}
    \psfrag{X32}{$X_{32}$}
    \psfrag{Y32}{$Y_{32}$}
    \psfrag{Z32}{$Z_{32}$}
    \psfrag{X14}{$X_{14}$}
    \psfrag{X43}{$X_{43}$}
    \psfrag{Y43}{$Y_{43}$}
    \psfrag{Q1a}{$Q_{1a}$}
    \psfrag{Qa2}{$Q_{a2}$}
    \psfrag{Q2b}{$Q_{2b}$}
    \psfrag{Qb3}{$Q_{b3}$}
    \psfrag{Q3c}{$Q_{3c}$}
    \psfrag{Qc4}{$Q_{c4}$}
    \psfrag{Q4d}{$Q_{4d}$}
    \psfrag{Qd1}{$Q_{d1}$}
    \includegraphics[scale=1.20]{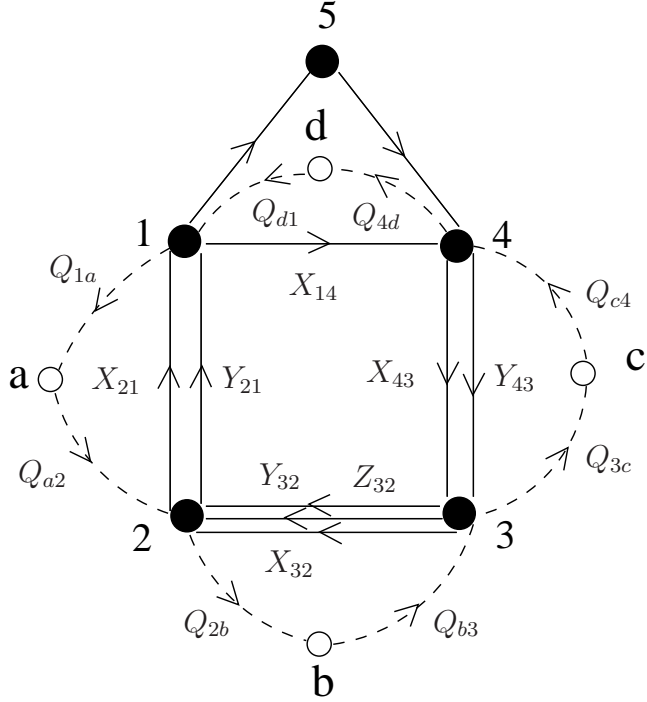}
    \caption{\small Quiver diagram with flavors. White nodes denote
      flavor groups.}
  \label{arbquiv1}
\end{figure}

\subsubsection{Seiberg duality and one-loop masses}

\begin{figure}[!htp]
    \centering
    \psfrag{X21}{$X_{21}$}
    \psfrag{Y21}{$Y_{21}$}
    \psfrag{X32}{$X_{32}$}
    \psfrag{Y32}{$Y_{32}$}
    \psfrag{Z32}{$Z_{32}$}
    \psfrag{Qa2}{$Q_{a2}$}
    \psfrag{Q2b}{$Q_{2b}$}
    \includegraphics[scale=1.20]{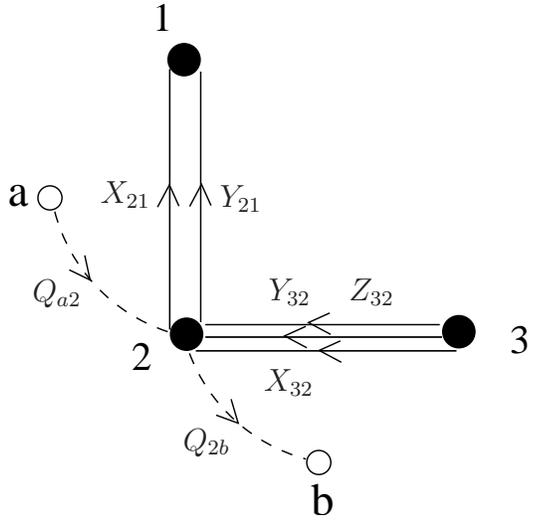}
  \caption{\small Relevant part of quiver before Seiberg duality.}
  \label{dualarbquiv1}
\end{figure}

\begin{figure}[!htp]
    \centering
    \psfrag{X12}{$\tilde{X}_{12}$}
    \psfrag{Y12}{$\tilde{Y}_{12}$}
    \psfrag{X23}{$\tilde{X}_{23}$}
    \psfrag{Y23}{$\tilde{Y}_{23}$}
    \psfrag{Z23}{$\tilde{Z}_{23}$}
    \psfrag{Qb2}{$\tilde{Q}_{b2}$}
    \psfrag{Q2a}{$\tilde{Q}_{2a}$}
    \psfrag{Xab}{$X_{ab}$}
    \psfrag{R1}{$R^1$}
    \psfrag{R2}{$R^2$}
    \psfrag{S1}{$S^1$}
    \psfrag{S2}{$S^2$}
    \psfrag{S3}{$S^3$}
    \psfrag{M1}{$M^1, \ldots , M^6$}    
    \includegraphics[scale=1.20]{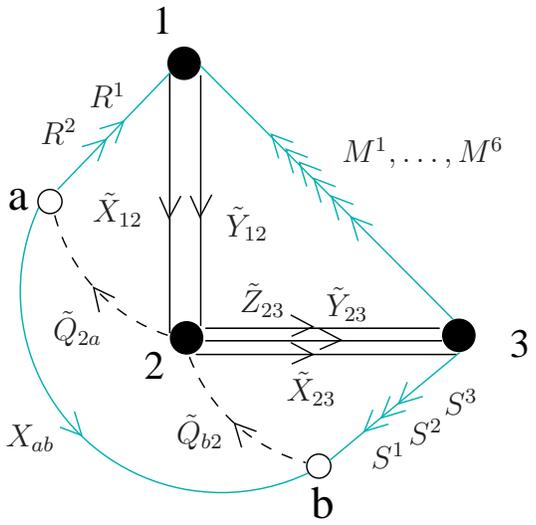}
  \caption{\small Relevant part of the quiver after Seiberg duality on node 2.}
  \label{dualarbquiv2}
\end{figure}

We consider introducing a number of massive flavors such that node 2
is in the free magnetic phase, and consider its Seiberg dual. The only
relevant fields in this case are those charged under gauge factor 2,
as shown if Figure \ref{dualarbquiv1}.
The Seiberg dual gives us Figure \ref{dualarbquiv2} 
where the $M$'s are mesons with indices in the gauge groups, $R$'s and
$S$'s are mesons with only one index in the flavor group, and $X_{ab}$
is a meson with both indices in the flavor groups. The original cubic
superpotential and flavor mass superpotentials become
\begin{eqnarray}
  W_{flavor\,dual} & = & \lambda' \,( \,
  S^1_{3b}Q_{b3} \, + \, R^1_{a1}Q_{1a} \, + \,
  X_{14}Q_{4d}Q_{d1}\, + \, Y_{43}Q_{3c}Q_{c4}  \,) \nonumber \\
  W_{mass\,dual} & = & m_2\underline{X_{ab}} \, + \, m_3 Q_{b3}Q_{3c} \, + \, m_4
  Q_{c4}Q_{4d}\, + \, m_1 Q_{d1}Q_{1a}
\end{eqnarray}
In addition we have the extra meson superpotential
\begin{eqnarray}
  W_{mesons} & = & h\,( \,  \underline{X_{ab}\tilde{Q}_{b2}\tilde{Q}_{2a}} \,  +  \, R^1_{a1}\tilde{X}_{12}\tilde{Q}_{2a} \, + \, R^2_{a1}\tilde{Y}_{12}\tilde{Q}_{2a} 
  \,  +  \, S^1_{3b}\tilde{Q}_{b2}\tilde{X}_{23} \, + \,
  S^2_{3b}\tilde{Q}_{b2}\tilde{Y}_{23} \nonumber \\
  & + & \, S^3_{3b}\tilde{Q}_{b2}\tilde{Z}_{23} \,
  +  \, M^1_{31}\tilde{X}_{12}\tilde{X}_{23} \,  + \,
  M^2_{31}\tilde{X}_{12}\tilde{Y}_{23}  \,  +  \,
  M^3_{31}\tilde{X}_{12}\tilde{Z}_{23} \nonumber \\
  & + & \, M^4_{31}\tilde{Y}_{12}\tilde{X}_{23} \,  +  \,
  M^5_{31}\tilde{Y}_{12}\tilde{Y}_{23}  \,  +  \,
  M^6_{31}\tilde{Y}_{12}\tilde{Z}_{23} \, ).
\end{eqnarray}
The crucial point is that we always obtain terms of the kind
underlined above, namely a piece of the superpotential reading $m_2
X_{ab} + h X_{ab}\tilde{Q}_{b2}\tilde{Q}_{2a}$. This leads to tree
level supersymmetry breaking by the rank condition, as announced.
Moreover the superpotential fits in the structure of the generalized
asymmetric O'Raifeartaigh model studied in appendix
\ref{sec:basic-superpotentials}, with $X_{ab}$, $\tilde{Q}_{b2}$,
$\tilde{Q}_{2a}$ corresponding to $X$, $\phi_1$, $\phi_2$
respectively. The multiplets $\tilde{Q}_{b2}$ and $\tilde{Q}_{2a}$ are
split at tree level, and $X_{ab}$ is massive at 1-loop.  From our
study of the generalized asymmetric case, any field which has a cubic
coupling to the supersymmetry breaking fields $\tilde{Q}_{b2}$ or
$\tilde{Q}_{2a}$ is one-loop massive as well. Using the general
structure of $W_{mesons}$, a little thought shows that all dual quarks
with no flavor index (e.g. $\tilde{X}$, $\tilde{Y}$) and all mesons
with one flavor index (e.g. $R$ or $S$) couple to the supersymmetry
breaking fields.

Thus they all get one-loop masses (with positive squared mass).
Finally, the flavors of other gauge factors (e.g. $Q_{b3}$) are
massive at tree level from $W_{mass}$.

The bottom line is that the only fields which do not get mass from
these interactions are the mesons with no flavor index, and the
bi-fundamentals which do not get dualized (uncharged under node 2).
All these fields are related to the theory in the absence of extra
flavors, so they can be already stabilized at tree-level from the
original superpotential. So, the criteria for a metastable vacua is
that the original theory, {\em in the absence of flavors} leads, after
dualization of the node with $N_f<N_c$, to masses for all these fields
(or more mildly that they correspond to directions stabilized by mass
terms, or perhaps higher order superpotential terms).

For example, if we apply this criteria to the $dP_2$ case studied
previously, the original superpotential for the fractional DSB brane
is
\begin{eqnarray}
W = -\lambda X_{53}Y_{31}X_{15}
\end{eqnarray}
so after dualization we get 
\begin{eqnarray}
W = -\lambda M_{13}Y_{31}
\end{eqnarray}
which makes these fields massive. Hence this fractional brane, after adding the D7-branes in the appropriate configuration, will generate a metastable vacua will all
moduli stabilized.

The argument is completely general, and leads to an enormous
simplification in the study of the theories. In the next section we
describe several examples.  A more rigorous and elaborate proof is
provided in the appendix where we take into account the matricial
structure, and show that all fields, except for Goldstone bosons, get
positive squared masses at tree-level or at one-loop.

\subsection{Additional examples}

\subsubsection{The $dP_3$ case}

\begin{figure}
  \epsfxsize = 5.5cm
  \centerline{\epsfbox{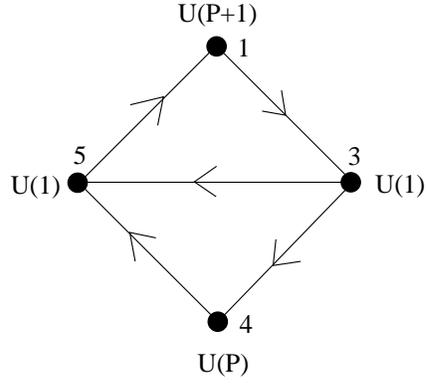}}
  \caption{\small Quiver diagram for the $dP_3$ theory with a DSB fractional brane.}
  \label{dp3}
\end{figure}

Let us consider the complex cone over $dP_3$, and introduce fractional
DSB branes of the kind considered in \cite{Franco:2005zu}. The quiver
is shown in Figure \ref{dp3} and the superpotential is
\begin{equation} 
W = X_{13}X_{35}X_{51} 
\end{equation} 
Node 1 has $N_f < N_c$ so upon addition of massive flavors and
dualization will lead to supersymmetry breaking by the rank condition.
Following the procedure of the previous section, we add $N_{f,1}$
flavors coupling to the bi-fundamentals $X_{13}$, $X_{35}$ and
$X_{51}$. Node 1 is in the free magnetic phase for $P + 1 \leq N_{f,1}
\, < \, \frac{3}{2}P+\frac 12$ .  Dualizing node 1, the above
superpotential becomes \begin{equation} W = X_{35}M_{53}
\end{equation} where $M_{53}$ is the meson $X_{51}X_{13}$. So,
following the results of the previous section, we can conclude that
this DSB fractional brane generates a metastable vacua with all
pseudomoduli lifted.

\subsubsection{Phase 1 of $PdP_4$}

\begin{figure}[!htp]
  \epsfxsize = 5.5cm
  \centerline{\epsfbox{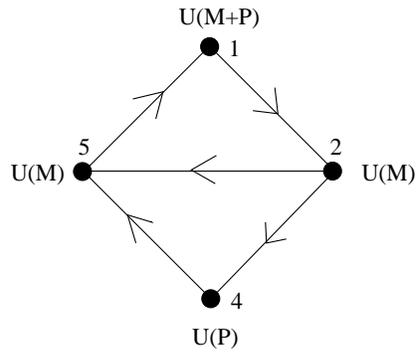}}
  \caption{\small Quiver diagram for the $dP_4$ theory with a DSB
    fractional branes.}
  \label{dp4}
\end{figure}

Let us consider the $PdP_4$ theory, and introduce the DSB fractional brane of the kind considered in \cite{Franco:2005zu}. The quiver is shown in
Figure \ref{dp4} .
The superpotential is 
\begin{equation} 
W = - X_{25}X_{51}X_{12} 
\end{equation} 
Node 1 has $N_f < N_c$ and will lead to supersymmetry breaking by the rank condition in the dual. Following the procedure of the previous section, we add $N_{f,1}$
flavors coupling to the bi-fundamentals $X_{12}$, $X_{25}$ and $X_{51}$. Node 1 is in the free magnetic phase for $P + 2 \leq  M + N_{f,1} \, < \, \frac{3}{2}(M+P)$.
Dualizing node 1, the above superpotential becomes $W =
X_{25}M_{52}$, where $M_{53}$ is the meson $X_{51}X_{12}$. Again we
conclude that this DSB fractional brane generates a metastable
vacua with all pseudomoduli lifted.

\subsubsection{The $Y^{p,q}$ family}

\begin{figure}[!htp]
  \epsfxsize = 2.5cm
  \centerline{\epsfbox{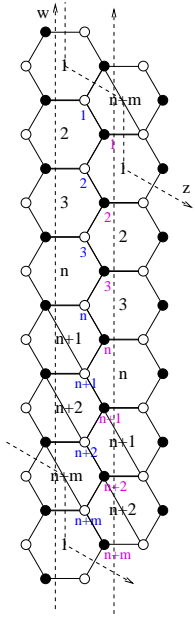}}
  \caption{The generic dimer for $Y^{p,q}$, from \cite{Brini:2006ej}.}
  \label{kast}
\end{figure}

Consider D3-branes at the real cones over the $Y^{p,q}$
Sasaki-Einstein manifolds
\cite{Gauntlett:2004zh,Gauntlett:2004yd,Gauntlett:2004hh,Martelli:2004wu},
whose field theory were determined in \cite{Benvenuti:2004dy}. The
theory admits a fractional brane \cite{Herzog:2004tr} of DSB kind,
which namely breaks supersymmetry and lead to runaway behavior
\cite{Franco:2005zu,Brini:2006ej}.  The analysis of metastability upon
addition of massive flavors for arbitrary $Y^{p,q}$'s is much more
involved than previous examples. Already the description of the field
theory on the fractional brane is complicated. Even for the simpler
cases of $Y^{p,q}$ and $Y^{p,p-1}$ the superpotential contains many
terms. In this section we do not provide a general proof of
metastability, but rather consider the more modest aim of showing that
all directions related to the runaway behavior in the absence of
flavors are stabilized by the addition of flavors. We expect that this
will guarantee full metastability, since the fields not involved in
our analysis parametrize directions orthogonal to the runaway at
infinity.
\\
The dimer for $Y^{p,q}$ is shown in Figure \ref{kast} and consists of
a column of $n$ hexagons and $2\,m$ quadrilaterals which are just
halved hexagons \cite{Brini:2006ej}. The labels $(n,m)$ are related to
$(p,q)$ by
\begin{equation} n= 2q \ \ ; \ \ m = p-q \end{equation}

\medskip

$\bullet$ {\bf The $Y^{p,1}$ case}

\begin{figure}[!htp]
  \epsfxsize = 2.0cm
  \centerline{\epsfbox{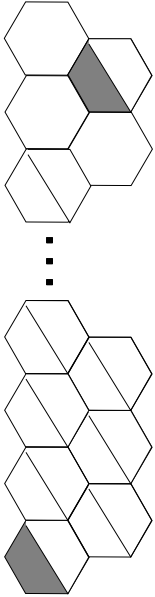}}
  \caption{The dimer for $Y^{p,1}$.}
  \label{yp1}
\end{figure}

\begin{figure}[!htp]
  \epsfxsize = 4.0cm
  \centerline{\epsfbox{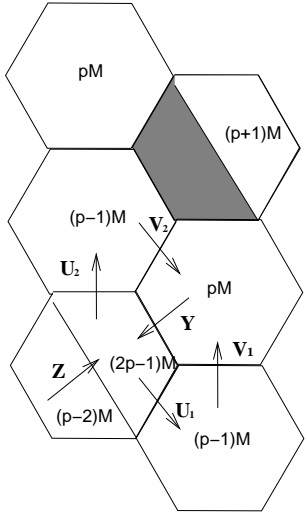}}
  \caption{Top part of the dimer for $Y^{p,1}$. The hexagons are labeled by the ranks of the respective gauge groups}
  \label{yp1b}
\end{figure}

The dimer for the theory on the DSB fractional brane in the $Y^{p,1}$
case is shown in Figure \ref{yp1}, a periodic array of a column of two
full hexagons, followed by $p-1$ cut hexagons (the shaded
quadrilateral has $N_c=0$). As shown in \cite{Brini:2006ej}, the top
quadrilateral which has $N_f<N_c$, and induces the ADS superpotential
triggering the runaway. The relevant part of the dimer is shown in
Figure \ref{yp1b}, where $V_1$ and $V_2$ are the fields that run to
infinity \cite{Brini:2006ej}.  This node will lead to supersymmetry
breaking by the rank condition in the dual. It is in the free magnetic
phase for $M + 1 \leq N_{f,1} \, < \, pM + \frac{M}{2}$.  The piece of
the superpotential involving the $V_1$ and $V_2$ terms is
\begin{equation}
W = Y U_2 V_2 - Y U_1 V_1.
\end{equation}
In the dual theory, the dual superpotential makes the fields massive. Hence, the theory has a metastable vacua where the runaway fields are stabilized.

\medskip

$\bullet$ {\bf The $Y^{p,p-1}$ case}

\begin{figure}[!htp]
  \epsfxsize = 4.0cm
  \centerline{\epsfbox{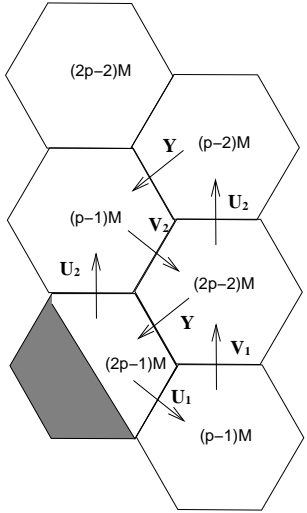}}
  \caption{Bottom part of the dimer for $Y^{p,p-1}$. The hexagons are labeled by the ranks of the respective gauge groups}
  \label{ypp1}
\end{figure}

The analysis for $Y^{p,p-1}$ is similar but in this case it is the
bottom quadrilateral which has the highest rank and thus gives the ADS
superpotential \cite{Brini:2006ej}. The relevant part of the dimer is
shown in Figure \ref{ypp1}, and the runaway direction is described by
the fields $V_1$ and $V_2$. Upon addition of $N_{f,1}$ flavors, the
relevant node in the in the free magnetic phase for $M + 1 \leq
N_{f,1} \, < \, pM + \frac{M}{2}$ Considering the superpotential, it
is straightforward to show that the runaway fields become massive.
Complementing this with our analysis in previous section, we conclude
that the theory has a metastable vacua where the runaway fields are
stabilized.

We have thus shown that we can obtain metastable vacua for fractional branes at cones over the $Y^{p,1}$ and $Y^{p,p-1}$ geometries. Although there is no obvious
generalization for arbitrary $Y^{p,q}$'s, our results strongly suggest that the existence of metastable vacua extends to the complete family.

\section{Conclusions and outlook}
\label{conclusions}

The present work introduces techniques and computations which suggest
that the existence of metastable supersymmetry breaking vacua is a
general property of quiver gauge theories on DSB fractional branes,
namely fractional branes associated to obstructed complex
deformations. It is very satisfactory to verify the correlation
between a non-trivial dynamical property in gauge theories and a
geometric property in their string theory realization. The existence
of such correlation fits nicely with the remarkable properties of
gauge theories on D-branes at singularities, and the gauge/gravity
correspondence for fractional branes.

Beyond the fact that our arguments do not constitute a general proof,
our analysis has left a number of interesting open questions. In fact,
as we have mentioned, all theories on DSB fractional branes contain
one or several fields which do not appear in the superpotential. We
expect the presence of these fields to have a direct physical
interpretation, which has not been uncovered hitherto. It would be
interesting to find a natural explanation for them.

Finally, a possible extension of our results concerns D-branes at
orientifold singularities, which can lead to supersymmetry breaking
and runaway as in \cite{Lykken:1998ec}. Interestingly, in this case
the field theory analysis is more challenging, since they would
require Seiberg dualities of gauge factors with matter in two-index
tensors. It is very possible that the string theory realization, and
the geometry of the singularity provide a much more powerful tool to
study the system.

Overall, we expect other surprises and interesting relations to come
up from further study of D-branes at singularities.

\section*{Acknowledgments}

We thank S. Franco for useful discussions.  A.U. thanks M. Gonz\'alez
for encouragement and support.  This work has been supported by the
European Commission under RTN European Programs MRTN-CT-2004-503369,
MRTN-CT-2004-005105, by the CICYT (Spain), and by the Comunidad de
Madrid under project HEPHACOS P-ESP-00346. The research by I.G.-E. is
supported by the Gobierno Vasco PhD fellowship program.  The research
of F.S is supported by the Ministerio de Educaci\'on y Ciencia through
an FPU grant. I.G.-E. and F.S. thank the CERN Theory Division for
hospitality during the completion of this work.

\newpage
\appendix

\section{Technical details about the calculation via Feynman diagrams}
\subsection{The basic amplitudes}
\label{basicamplitudes}
\begin{figure}[!htp]
  \begin{center}
  \input{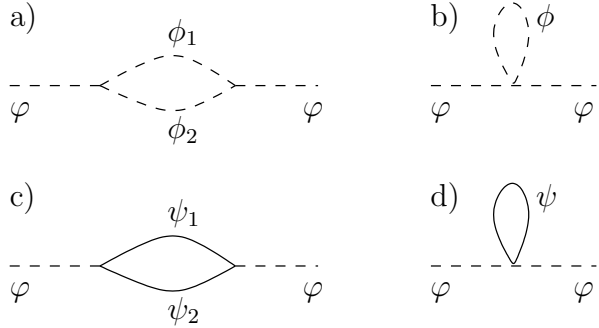}
  \end{center}
  \caption{\small Feynman diagrams contributing to the one-loop two
    point function. The dashed line denotes bosons and the solid one
    fermions.}
  \label{feynman_diagrams}
\end{figure}

In the main text we are interested in computing two point functions for the
pseudomoduli at one loop, and in section \ref{sec:goldstones} also
tadpole diagrams. There are just a few kinds of diagrams entering in
the calculation, which we will present now for the two-point
function, see Figure \ref{feynman_diagrams}. The (real) bosonic fields are denoted by $\phi_i$ and the
(Weyl) fermions by $\psi_i$. The pseudomodulus we are interested in is
denoted by $\varphi$.

\subsubsection*{Bosonic contributions}
These come from two terms in the Lagrangian. First there is a diagram
coming from terms of the form (Figure \ref{feynman_diagrams}b):
\begin{equation}
\cl = \ldots + \lambda \varphi^2 \phi^2 - \frac{1}{2}m^2\phi^2,
\end{equation}
giving an amplitude (we will be using dimensional regularization)
\begin{equation}
i\cam = \frac{-2i\lambda}{(4\pi)^2}m^2\left(
\frac{1}{\epsilon} - \g + 1 + \log4\pi -\log m^2\right).
\end{equation}
The other contribution comes from the diagram in Figure
\ref{feynman_diagrams}a:
\begin{equation}
\cl = \ldots + \lambda \varphi \phi_1 \phi_2 -
\frac{1}{2} m_1^2 \phi_1^2 - \frac{1}{2} m_2^2 \phi_2^2, 
\end{equation}
which contributes to the two point function with an amplitude:
\begin{equation}
i\cam = \frac{i\lambda^2}{(4\pi)^2}\left(
\frac{1}{\epsilon} - \g + \log4\pi - \int_0^1 dx \log \Delta
\right),
\end{equation}
where here and in the following we denote $\Delta \equiv x m_1^2 +
(1-x)m_2^2$.

\subsubsection*{Fermionic contributions}
The relevant vertices here are again of two possible kinds, one of
which is nonrenormalizable. The cubic interaction comes from terms in
the Lagrangian given by the diagram in Figure \ref{feynman_diagrams}c:
\begin{equation}
\cl = \ldots + \varphi(a\psi_1 \psi_2 + a^* \bar\psi_1\bar\psi_2) + 
\frac{1}{2} m_1 (\psi_1^2+\bar\psi_1^2) + \frac{1}{2} m_2
(\psi_2^2+\bar\psi_2^2).
\end{equation}
We are assuming real masses for the fermions here, in the
configurations we study this can always be achieved by an appropriate
field redefinition. The contribution from such vertices is given by:
\begin{eqnarray}
\nonumber i\cam & = & \int_0^1 dx \left\lbrace
\frac{-2i m_1m_2}{(4\pi)^2}(a^2 + (a^2)^*)\left(
\frac{1}{\epsilon} - \g + \log4\pi - \log \Delta
\right) \right.\\
&& - \left. \frac{8i|a|^2}{(4\pi)^2}\Delta\left(
\frac{1}{\epsilon} - \g + \log4\pi +\frac{1}{2} - \log \Delta
\right)
\right\rbrace.
\end{eqnarray}

The other fermionic contribution, which one does not need as long as
one is dealing with renormalizable interactions only (but we will need
in the main text when analyzing the pseudomodulus $\theta$), is given
by terms in the Lagrangian of the form (Figure \ref{feynman_diagrams}d):
\begin{equation}
\cl = \ldots + \lambda \varphi^2 (\psi^2 + \bar\psi^2) +
\frac{1}{2} m (\psi^2 + \bar\psi^2),
\end{equation}
which contributes to the total amplitude with:
\begin{equation}
i\cam = \frac{8\lambda m i}{(4\pi)^2} m^2 \left(
\frac{1}{\epsilon} - \g + 1 + \log4\pi - \log m^2
\right).
\end{equation}

\subsection{The basic superpotentials}
\label{sec:basic-superpotentials}
The previous amplitudes are the basic ingredients entering the
computation, but in general the number of diagrams contributing to the
two point amplitudes is quite big, so calculating all the
contributions by hand can get quite involved in particular
examples\footnote{The authors wrote the computer program in {\tt
    http://cern.ch/inaki/pm.tar.gz } which helped greatly in the
  process of computing the given amplitudes for the relevant models.}.
Happily, one finds that complicated models (such as $dP_1$ or $dP_2$,
studied in the main text) reduce to performing the analysis for only
two different superpotentials, which we analyze in this section.

\subsubsection*{The symmetric case}
We want to study in this section a superpotential of the form:
\begin{equation}
W = h(X\phi_1\phi_2 + \mu \phi_1 \phi_3 + \mu \phi_2 \phi_4 - \mu^2 X).
\end{equation}
This model is a close cousin of the basic O'Raifeartaigh model. We are
interested in the one loop contribution to the two point function of
$X$, which is massless at tree level.

From the (F-term) bosonic potential one obtains the following terms
entering the one loop computation:
\begin{eqnarray}
\nonumber V & = & \left[
|h X \phi_2|^2 + |h|^2 \mu(X\phi_2\phi_3^* + X^* \phi_2^* \phi_3) +
|h|^2 \mu(X\phi_1\phi_4^* + X^* \phi_1^* \phi_4)\right] \\
&+& |h|^2 \mu^2(\phi_1\phi_2 + \phi_1^*\phi_2^*) +
\sum_{i=1}^4 |h|^2 \mu^2 |\phi_i|^2
\end{eqnarray}

In order to do the computation it is useful to diagonalize the mass
matrix by introducing $\phi_+$ and $\phi_-$ such that:
\begin{equation}
\phi_1 = \frac{1}{\sqrt{2}}(\phi_+ + i\phi_-) \qquad
\phi_2 = \frac{1}{\sqrt{2}}(\phi_+ - i\phi_-)
\end{equation}
and $\phi_a$, $\phi_b$ such that:
\begin{equation}
\phi_3^* = \frac{1}{\sqrt{2}}(\phi_a + i\phi_b) \qquad
\phi_4^* = \frac{1}{\sqrt{2}}(\phi_a - i\phi_b).
\end{equation}
With these redefinitions the bosonic scalar potential decouples into
identical $\phi_+$ and $\phi_-$ sectors, giving two decoupled copies
of:
\begin{eqnarray}
\nonumber V & = & |h|^2 |X|^2 |\phi_+|^2 + |h|^2\mu^2 (|\phi_+|^2 +
|\phi_a|^2) \\
&& +|h|^2\mu (X\phi_+\phi_a + X^*\phi_+^*\phi_a^*) -
\frac{|h|^2\mu^2}{2}\left(\phi_+^2 + (\phi_+^2)^*\right).
\end{eqnarray}
Calculating the amplitude consists simply of constructing the (very
few) two point diagrams from the potential above and plugging the
formulas above for each diagram (the fermionic part is even simpler in
this case). The final answer is that in this model the one loop
correction to the mass squared of $X$ is given by:
\begin{equation}
\delta m_X^2 = \frac{|h^4|\mu^2}{8\pi^2}(\log 4 - 1).
\end{equation}

\subsubsection*{The generalized asymmetric case}
The next case is slightly more complicated, but will suffice to
analyze completely all the models we encounter. We will be interested
in the one loop contribution to the mass of the pseudomoduli $Y$ in a
theory with superpotential:
\begin{equation}
W = h(X\phi_1\phi_2 + \mu \phi_1 \phi_3 + \mu \phi_2 \phi_4 - \mu^2 X)
+ k(r Y\phi_1\phi_5 + \mu\phi_5\phi_7),
\end{equation}
with $k$ and $r$ arbitrary complex numbers. The procedure is
straightforward as above, so we will just quote the result. We obtain
an amplitude given by:
\begin{equation}
i\cam = \frac{-i}{(4\pi)^2} |h^2 r \mu|^2
\cc\left(\frac{|k|^2}{|h|^2}\right),
\end{equation}
where we have defined $\cc(t)$ as:
\begin{equation}
\cc(t) = \frac{t}{2-t}\left( \log 4 - \frac{t}{t-1}\log t \right).
\end{equation}
Note that this is a positive definite function, meaning that the one
loop correction to the mass is always positive, and the pseudomoduli
get stabilized for any (nonzero) value of the parameters. Also note
that the limit of vanishing $t$ with $|r|^2t$ fixed (i.e., vanishing
masses for $\phi_5$ and $\phi_7$, but nonvanishing coupling of $Y$ to
the supersymmetry breaking sector) gives a nonvanishing contribution
to the mass of $Y$.

\section{D7--branes in the Riemann surface}
\label{Riemannsurface}

\begin{figure}[!htp]
    \centering
    \includegraphics[scale=0.5]{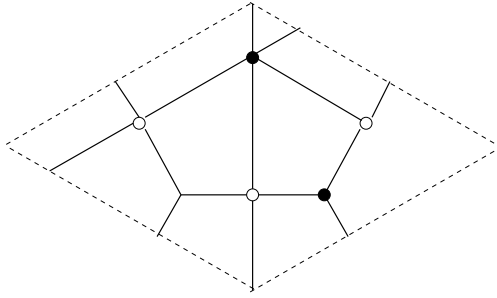}
  \caption{\small Dimer diagram for D3--branes at a $dP_2$ singularity.}
  \label{dp2_dimer3}
\end{figure}

\begin{figure}[!htp]
    \centering
    \includegraphics[scale=0.5]{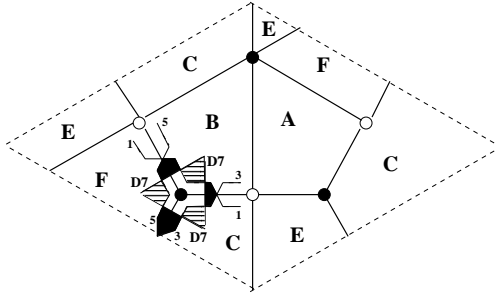}
    \caption{\small Riemann surface in the geometry mirror to the
      complex cone over $dP_2$, shown as a tiling of a $T^2$ with
      punctures (denoted by capital letters). The figure shows the
      non-compact 1-cycles extending between punctures, corresponding
      to D7-branes, and a piece of the 1-cycles that correspond to the
      mirror of the D3-branes.}
  \label{dp2_dimer2}
\end{figure}

The gauge theory of D3-branes at toric singularities can be encoded in
a dimer diagram
\cite{Hanany:2005ve,Franco:2005rj,Hanany:2005ss,Feng:2005gw,Franco:2006gc}.
This corresponds to a bi-partite tiling of $T^2$, where faces
correspond to gauge groups, edges correspond to bi-fundamentals, and
nodes correspond to superpotential terms. As an example, the dimer
diagram of D3--branes on the cone over $dP_2$ is shown in Figure
\ref{dp2_dimer3}. As shown in \cite{Feng:2005gw}, D3--branes on a
toric singularity are mirror to D6--branes on intersecting 3-cycles in
a geometry given by a fibration of a Riemann surface $\Sigma$ with
punctures. This Riemann surface is just a thickening of the web
diagram of the toric singularity
\cite{Aharony:1997ju,Aharony:1997bh,Leung:1997tw}, with punctures
associated to external legs of the web diagram. The mirror D6-branes
wrap non-trivial 1-cycles on this Riemann surface, with their
intersections giving rise to bi-fundamental chiral multiplets, and
superpotential terms arising from closed discs bounded by the
D6-branes. In \cite{Franco:2006es}, it was shown that D7--branes
passing through the singular point can be described in the mirror
Riemann surface $\Sigma$ by non-compact 1-cycles which come from
infinity at one puncture and go to infinity at another. Figure
\ref{dp2_dimer2} shows the 1-cycles corresponding to some D3- and
D7-branes in the Riemann surface in the geometry mirror to the complex
cone over $dP_2$. A D7-brane leads to flavors for the two D3-brane
gauge factors whose 1-cycles are intersected by the D7-brane 1-cycle,
and there is a cubic coupling among the three fields (related to the
disk bounded by the three 1-cycles in the Riemann surface).

\begin{figure}[!htp]
    \centering
    \psfrag{Q1i}{$Q_{1i}$}
    \psfrag{Qi3}{$Q_{i3}$}
    \psfrag{Q3j}{$Q_{3j}$}
    \psfrag{Qj5}{$Q_{j5}$}
    \psfrag{Q5k}{$Q_{5k}$}
    \psfrag{Qk1}{$Q_{k1}$}
    \includegraphics[scale=2]{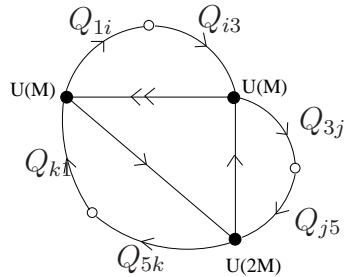}
  \caption{\small Quiver for the $dP_2$ theory with M fractional branes and flavors.}
  \label{dp2_flav2}
\end{figure}

As stated in Section \ref{generalcase}, given a gauge theory of
D3-branes at a toric singularity, we introduce flavors for some of the
gauge factors in a specific way. We pick a term in the superpotential,
and we introduce flavors for all the involved gauge factors, and
coupling to all the involved bifundamental multiplets. For example,
the quiver with flavors for the $dP_2$ theory is shown in Figure
\ref{dp2_flav2}.

On the Riemann surface, this procedure amounts to picking a node and introducing D7-branes crossing all the edges ending on the node, see Figure \ref{dp2_dimer2}.
In this example we obtain the superpotential terms
\begin{eqnarray} 
W_{flavor} =
\lambda'(Q_{1i}\tilde{Q}_{i3}Y_{31}+Q_{3j}\tilde{Q}_{j5}X_{53}+
Q_{5k}\tilde{Q}_{k1}X_{15})
\label{W_dP2_flavor2}
\end{eqnarray}
In addition we introduce mass terms 
\begin{eqnarray} W_{mass} = m_{1}Q_{1i}\tilde{Q}_{k1} +
m_{2}Q_{3j}\tilde{Q}_{i3} + m_{5}Q_{5k}\tilde{Q}_{j5}
\label{W_dP2_mass2}
\end{eqnarray}
This procedure is completely general and applies to all gauge theories for branes at toric singularities\footnote{This  procedure does not apply if the superpotential (regarded as a loop in the quiver) passes twice through the node which is eventually dualized in the derivation of the metastable vacua. However we have found no example of this for any DSB fractional branes.}.

\section{Detailed proof of Section \ref{generalcase}}
\label{PROOF}
\begin{figure}[!htp]
    \centering
    \psfrag{X21}{$X_{21}$}
    \psfrag{Y21}{$Y_{21}$}
    \psfrag{X32}{$X_{32}$}
    \psfrag{Y32}{$Y_{32}$}
    \psfrag{Z32}{$Z_{32}$}
    \psfrag{X14}{$X_{14}$}
    \psfrag{X43}{$X_{43}$}
    \psfrag{Y43}{$Y_{43}$}
    \psfrag{Q1a}{$Q_{1a}$}
    \psfrag{Qa2}{$Q_{a2}$}
    \psfrag{Q2b}{$Q_{2b}$}
    \psfrag{Qb3}{$Q_{b3}$}
    \psfrag{Q3c}{$Q_{3c}$}
    \psfrag{Qc4}{$Q_{c4}$}
    \psfrag{Q4d}{$Q_{4d}$}
    \psfrag{Qd1}{$Q_{d1}$}
    \includegraphics[scale=0.9]{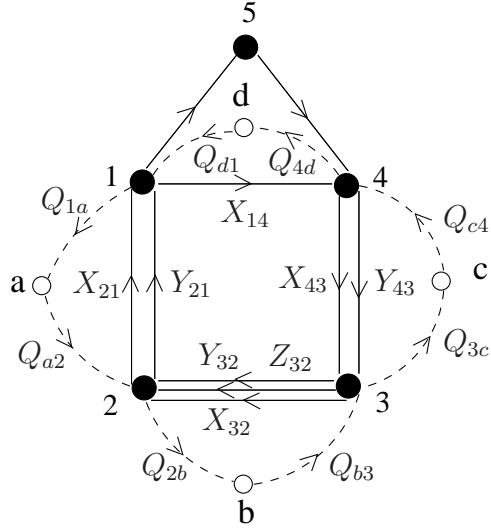}
    \caption{\small Quiver diagram with flavors. White nodes denote flavor groups}
  \label{arbquiv11}
\end{figure}

\begin{figure}[!htp]
    \centering
    \psfrag{X21}{$X_{21}$}
    \psfrag{Y21}{$Y_{21}$}
    \psfrag{X32}{$X_{32}$}
    \psfrag{Y32}{$Y_{32}$}
    \psfrag{Z32}{$Z_{32}$}
    \psfrag{Qa2}{$Q_{a2}$}
    \psfrag{Q2b}{$Q_{2b}$}
    \includegraphics[scale=0.9]{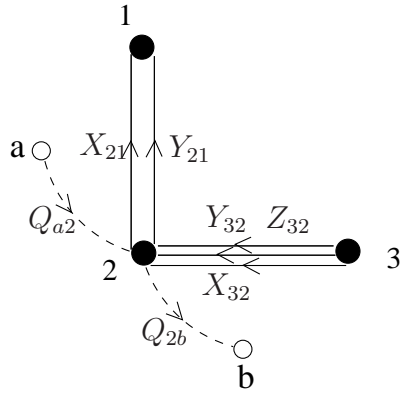}
  \caption{\small Relevant part of quiver before Seiberg duality.}
  \label{dualarbquiv11}
\end{figure}

\begin{figure}[!htp]
    \centering
    \psfrag{X12}{$\tilde{X}_{12}$}
    \psfrag{Y12}{$\tilde{Y}_{12}$}
    \psfrag{X23}{$\tilde{X}_{23}$}
    \psfrag{Y23}{$\tilde{Y}_{23}$}
    \psfrag{Z23}{$\tilde{Z}_{23}$}
    \psfrag{Qb2}{$\tilde{Q}_{b2}$}
    \psfrag{Q2a}{$\tilde{Q}_{2a}$}
    \psfrag{Xab}{$X_{ab}$}
    \psfrag{R1}{$R^1$}
    \psfrag{R2}{$R^2$}
    \psfrag{S1}{$S^1$}
    \psfrag{S2}{$S^2$}
    \psfrag{S3}{$S^3$}
    \psfrag{M1}{$M^1, \ldots , M^6$}    
    \includegraphics[scale=0.9]{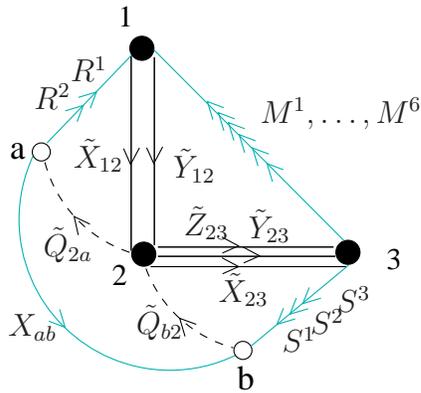}
  \caption{\small Relevant part of the quiver after Seiberg duality on node 2.}
  \label{dualarbquiv21}
\end{figure}

Recall that in Section \ref{generalcase} we considered the
illustrative example of the gauge theory given by the quiver in Figure
\ref{arbquiv11}. Since node 2 is the one we wish to dualize, the only
relevant part of the diagram is shown in Figure \ref{dualarbquiv11}.
We show the Seiberg dual in Figure \ref{dualarbquiv21}.
The above choice of D7--branes, which we showed in appendix
\ref{Riemannsurface} can be applied to arbitrary toric singularities,
gives us the superpotential terms
\begin{eqnarray}
  W_{flavor} & = & \lambda' \,(
  \, X_{32}Q_{2b}Q_{b3} \, + \, X_{21}Q_{1a}Q_{a2} \, + \,
  X_{14}Q_{4d}Q_{d1}\, + \, Y_{43}Q_{3c}Q_{c4}  \,) \nonumber \\
  W_{mass} & = & m_2 Q_{a2}Q_{2b} \, + \, m_3 Q_{b3}Q_{3c} \, + \, m_4
  Q_{c4}Q_{4d}\, + \, m_1 Q_{d1}Q_{1a}
\end{eqnarray}
Taking the Seiberg dual of
node 2 gives
\begin{eqnarray}
  W_{flavor\,dual} & = & \lambda' \,( \,
  S^1_{3b}Q_{b3} \, + \, R^1_{a1}Q_{1a} \, + \,
  X_{14}Q_{4d}Q_{d1}\, + \, Y_{43}Q_{3c}Q_{c4}  \,) \nonumber \\
  W_{mass\,dual} & = & m_2\underline{X_{ab}} \, + \, m_3 Q_{b3}Q_{3c}
  \, + \, m_4
  Q_{c4}Q_{4d}\, + \, m_1 Q_{d1}Q_{1a} \nonumber\\
  W_{mesons} & = & h\,( \,
  \underline{X_{ab}\tilde{Q}_{b2}\tilde{Q}_{2a}} \nonumber \\
  \, & + & \, R^1_{a1}\tilde{X}_{12}\tilde{Q}_{2a} \, + \, R^2_{a1}\tilde{Y}_{12}\tilde{Q}_{2a} \nonumber \\
  \, & + & \, S^1_{3b}\tilde{Q}_{b2}\tilde{X}_{23} \, + \, S^2_{3b}\tilde{Q}_{b2}\tilde{Y}_{23} \, + \,  
  S^3_{3b}\tilde{Q}_{b2}\tilde{Z}_{23} \nonumber \\
  \, & + & \, M^1_{31}\tilde{X}_{12}\tilde{X}_{23} \,  + \, M^2_{31}\tilde{X}_{12}\tilde{Y}_{23}  \,  +  \,  M^3_{31}\tilde{X}_{12}\tilde{Z}_{23}
  \nonumber \\
  \, & + & \, M^4_{31}\tilde{Y}_{12}\tilde{X}_{23} \,  +  \,
  M^5_{31}\tilde{Y}_{12}\tilde{Y}_{23}  \,  +  \,
  M^6_{31}\tilde{Y}_{12}\tilde{Z}_{23} \, )
  \label{supos}
\end{eqnarray}
where we have not included the original superpotential.  The
crucial point is that the underlined terms appear for any quiver
gauge theory with flavors introduced as described in appendix \ref{Riemannsurface}. 
As described in the main text, supersymmetry is broken by the rank condition due to the F-term of the dual meson associated to the massive flavors. Our vacuum ansatz is 
(we take $N_f = 2$ and $N_c = 1$ for simplicity; this does not affect our conclusions)
\begin{equation}
\begin{array}{cc}
  \tilde{Q}_{b2}\, =\, \pmatrix{\mu \mathbf{1}_{N_c} \cr 0}  & \, ; \   \tilde{Q}_{2a}\, =\, (\mu \mathbf{1}_{N_c} \, ; \,  0) 
\end{array}
\label{vev2}
\end{equation} 
with all other vevs set to zero. We parametrize the perturbations around this minimum as
\begin{equation}
\begin{array}{ccc}
  \tilde{Q}_{b2}\, =\, \pmatrix{\mu + \phi_1 \cr \phi_2 }   & \, ; \ {\tilde Q}_{2a}\, =\, ( \mu + \phi_3 \, ; \, \phi_4)   \,  & ; \ X_{ab} =\, \pmatrix{ X_{00} & X_{01} \cr X_{10} & X_{11}} 
\end{array}
\label{pert2}
\end{equation}
and the underlined terms give
\begin{eqnarray} h
X_{ab}\tilde{Q}_{b2}\tilde{Q}_{2a} - h \mu^2 X_{ab} & = & h X_{11}\,
\phi_2 \,\phi_4 - h \mu^2 X_{11} + h \mu \, \phi_2 \, X_{01} + h \mu
\, \phi_4 \, X_{10}
\nonumber \\
& + & h \mu \, \phi_1 \, X_{00} + h \mu \, \phi_3 \, X_{00} + h \,
\phi_1 \, \phi_3 X_{00} + h \, \phi_2 \, \phi_3 X_{01}
\nonumber \\
& + & h \, \phi_1 \, \phi_4 X_{10}
\label{W_oraf}
\end{eqnarray}
It is important to note that all the fields in (\ref{pert2}) will have
quadratic couplings only in  the underlined term
(\ref{W_oraf}). Thus, one can safely study this term, and the conclusions are independent of the other terms in the superpotential. Diagonalizing (\ref{W_oraf}) gives
\begin{eqnarray} h
X_{ab}\tilde{Q}_{b2}\tilde{Q}_{2a} - h \mu^2 X_{ab} & = & h X_{11}\,
\phi_2 \,\phi_4 - h \mu^2 X_{11} + h \mu \, \phi_2 \, X_{01} + h \mu
\, \phi_4 \, X_{10}
\nonumber \\
& + & \sqrt{2} h \mu \,  \phi_+ \, X_{00}  + \frac{h}{2} \,
\phi_+^2 \,  X_{0 0} - \frac{h}{2} \,
\phi_-^2 \,  X_{0 0} 
\nonumber \\
& + &  \frac{h}{\sqrt{2}}\,(\xi_+ - \xi_-)  \, \phi_2 X_{01} + \frac{h}{\sqrt{2}}\,(\xi_+ + \xi_-)  \, \phi_4 X_{10}
\label{W_oraf1}
\end{eqnarray}
where
\begin{equation}
\xi_+ = \frac{1}{\sqrt{2}}\,(\phi_1 + \phi_3) \ \ \ \  ; \ \ \ \  \xi_- = \frac{1}{\sqrt{2}}\,(\phi_1 - \phi_3) 
\end{equation}
This term is similar to the generalized asymmetric case studied in appendix \ref{sec:basic-superpotentials} with
\begin{equation}
X_{11} \rightarrow X \, ; \, \phi_4  \rightarrow \phi_1 \, ; \, \phi_2  \rightarrow \phi_2 \, ; \, X_{10}  \rightarrow \phi_3 \, ; \, X_{01}  \rightarrow \phi_4 \, 
\end{equation}
So here $X_{11}$ is the linear term that breaks supersymmetry, and $\phi_2$,
$\phi_4$ are the broken supersymmetry fields.  In (\ref{W_oraf1}), the only
massless fields at tree-level are $X_{11}$ and $\xi_-$. Comparing to
the ISS case in Section \ref{secISS} shows that $\im\xi_-$ is a
Goldstone boson and $X_{11}$, $\re \xi_-$ get mass at tree-level. As
for $\phi_2$ and $\phi_4$, setting $\rho_+ = \frac{1}{\sqrt{2}}(\phi_2
+ \phi_4)$ and $\rho_- = \frac{1}{\sqrt{2}}(\phi_2 - \phi_4)$ gives us
Re$(\rho_+)$ and $\im(\rho_-)$ massless and the rest massive.
Following the discussion in Section \ref{secISS}, Re$(\rho_+)$ and
$\im(\rho_-)$ are just the Goldstone bosons of the broken $SU(N_f)$
symmetry\footnote{In the case where the flavor group is $SU(2)$, these
  Goldstone bosons are associated to the generators $t_x$ and $t_y$.}.
We have thus shown that the dualized flavors (e.g. $\tilde{Q}_{b2}$,
$\tilde{Q}_{2a}$) and the meson with two flavor indices (e.g.
$X_{ab}$) get mass at tree-level or at 1-loop unless they are
Goldstone bosons.  Now, we need to verify that this is the case for
 the remaining fields.

\begin{figure}[!htp]
    \centering
    \psfrag{X14}{$X_{14}$}
    \psfrag{X43}{$X_{43}$}
    \psfrag{Y43}{$Y_{43}$}
    \psfrag{Q1a}{$Q_{1a}$}
    \psfrag{Qb3}{$Q_{b3}$}
    \psfrag{Q3c}{$Q_{3c}$}
    \psfrag{Qc4}{$Q_{c4}$}
    \psfrag{Q4d}{$Q_{4d}$}
    \psfrag{Qd1}{$Q_{d1}$}
    \psfrag{X12}{$\tilde{X}_{12}$}
    \psfrag{Y12}{$\tilde{Y}_{12}$}
    \psfrag{X23}{$\tilde{X}_{23}$}
    \psfrag{Y23}{$\tilde{Y}_{23}$}
    \psfrag{Z23}{$\tilde{Z}_{23}$}
    \psfrag{Qb2}{$\tilde{Q}_{b2}$}
    \psfrag{Q2a}{$\tilde{Q}_{2a}$}
    \psfrag{Xab}{$X_{ab}$}
    \psfrag{R1}{$R^1$}
    \psfrag{R2}{$R^2$}
    \psfrag{S1}{$S^1$}
    \psfrag{S2}{$S^2$}
    \psfrag{S3}{$S^3$}
    \psfrag{M1}{$M^1..M^6$}    
    \includegraphics[scale=0.8]{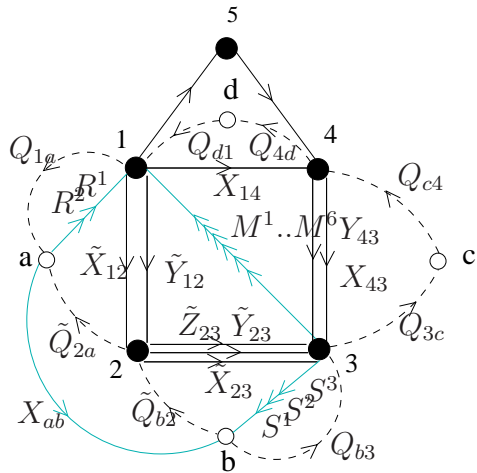}
  \caption{\small Quiver after Seiberg duality on node 2.}
  \label{dualarbquiv3}
\end{figure}
\begin{figure}[!htp]
    \centering
    \psfrag{X12}{$\tilde{X}_{12}$}
    \psfrag{Y12}{$\tilde{Y}_{12}$}
    \psfrag{X23}{$\tilde{X}_{23}$}
    \psfrag{Y23}{$\tilde{Y}_{23}$}
    \psfrag{Z23}{$\tilde{Z}_{23}$}
    \psfrag{Qb2}{$\tilde{Q}_{b2}$}
    \psfrag{Q2a}{$\tilde{Q}_{2a}$}
    \psfrag{Xab}{$X_{ab}$}
    \psfrag{R1}{$R^1$}
    \psfrag{R2}{$R^2$}
    \psfrag{S1}{$S^1$}
    \psfrag{S2}{$S^2$}
    \psfrag{S3}{$S^3$}
    \psfrag{M1}{$M^1, \ldots , M^6$}    
    \psfrag{Q1a}{$Q_{1a}$}
    \psfrag{Qb3}{$Q_{b3}$}
    \psfrag{Q3c}{$Q_{3c}$}
    \psfrag{Qd1}{$Q_{d1}$}
    \includegraphics[scale=0.8]{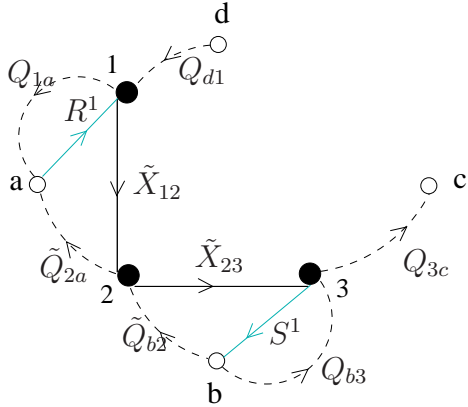}

  \caption{\small Relevant part of dual quiver for first class of bi-fundamentals.}
  \label{dualarbquiv4}
\end{figure}

The Seiberg dual of the original quiver diagram is shown in Figure \ref{dualarbquiv3}.
The dualized bi-fundamentals come in two classes. The first are
the ones that initially (before dualizing) had cubic flavor couplings,
there will always be only two of those (e.g. $\tilde{X}_{12}$,
$\tilde{X}_{23}$). The second are those that did not initially have
cubic couplings to flavors, there is an arbitrary number of those
(e.g. $\tilde{Y}_{12}$, $\tilde{Y}_{23}$, $\tilde{Z}_{23}$).  Figure
\ref{dualarbquiv4} shows the relevant part of the quiver for the first
class.
Recalling the superpotential terms (\ref{supos}), there are several possible sources of tree-level masses. For instance, these can arise in
$W_{flavor\,dual}$ and $W_{mass\,dual}$. Also, remembering our
assignation of vevs in (\ref{vev2}), tree-level masses can also arise
in $W_{mesons}$ from cubic couplings involving the broken supersymmetry 
fields (e.g. $\tilde{Q}_{b2}$, $\tilde{Q}_{2a}$). The first class of
bi-fundamentals (e.g. $\tilde{X}_{12}$, $\tilde{X}_{23}$) only appear
in $W_{mesons}$ coupled to their respective mesons (e.g. $R^1$,
$S^1$). In turn these mesons will appear in quadratic terms in
$W_{flavor\,dual}$ coupled to flavors (e.g. $S^1_{3b}Q_{b3}$ and
$R^1_{a1}Q_{1a}$), and these flavors each appear in one term in
$W_{mass}$.  Thus there
are two sets of three terms which are coupled at tree-level and which
always couple in the same way.  Consider for instance the term
\begin{eqnarray}
\label{eq:general-mass-mixing}
\lambda' \,S^1_{3b}Q_{b3} \,  + \, m_3 Q_{b3}Q_{3c} \, +  \,
h\, S^1_{3b}\tilde{Q}_{b2}\tilde{X}_{23} 
& = & \lambda' \, (S_1 \,  \, S_2) \pmatrix{B_1 \cr B_2} + m_1 (C_1 \,  \, C_2) \pmatrix{B_1 \cr B_2}
\nonumber \\ 
& +& \, h\, (S_1 \,  \, S_2) \pmatrix{\mu + \phi_1 \cr \phi_2 } \tilde{X}_{23}
\nonumber \\
& = & \lambda' (S_1 B_1 +  S_2 B_2) + m_1 (B_1 C_1 +  B_2 C_2) 
\nonumber \\
& + & h \mu \, S_1 \,\tilde{X}_{23}
\,+ \,h \,S_1 \, \phi_1 \,\tilde{X}_{23} \,+ \,h\, S_2\,  \phi_2\, \tilde{X}_{23}
\nonumber \\
\end{eqnarray}
where $S_i$, $B_i$, $C_i$ and $\tilde{X}_{23}$ are the perturbations
around the minimum. Diagonalizing (which can be done analytically for any
values of the couplings), we get that all terms except one get
tree-level masses, the massless field being:
\begin{equation}
  Y = m_1 S_2 - \lambda'C_2
\end{equation}
This massless field has a cubic coupling to $\phi_2\, \tilde{X}_{23}$
and gets mass at 1-loop since $\phi_2$ is a broken supersymmetry field, as
described in appendix \ref{sec:basic-superpotentials}.

Figure \ref{dualarbquiv5} shows the relevant part of the quiver for
the second class of bi-fundamentals (i.e. those that are dualized but
do not have cubic flavor couplings).
\begin{figure}[!htp]
    \centering
    \psfrag{X12}{$\tilde{X}_{12}$}
    \psfrag{Y12}{$\tilde{Y}_{12}$}
    \psfrag{X23}{$\tilde{X}_{23}$}
    \psfrag{Y23}{$\tilde{Y}_{23}$}
    \psfrag{Z23}{$\tilde{Z}_{23}$}
    \psfrag{Qb2}{$\tilde{Q}_{b2}$}
    \psfrag{Q2a}{$\tilde{Q}_{2a}$}
    \psfrag{Xab}{$X_{ab}$}
    \psfrag{R1}{$R^1$}
    \psfrag{R2}{$R^2$}
    \psfrag{S1}{$S^1$}
    \psfrag{S2}{$S^2$}
    \psfrag{S3}{$S^3$}
    \psfrag{M1}{$M^1, \ldots , M^6$}    
    \psfrag{Qc4}{$Q_{c4}$}
    \psfrag{Q4d}{$Q_{4d}$}
    \includegraphics[scale=0.8]{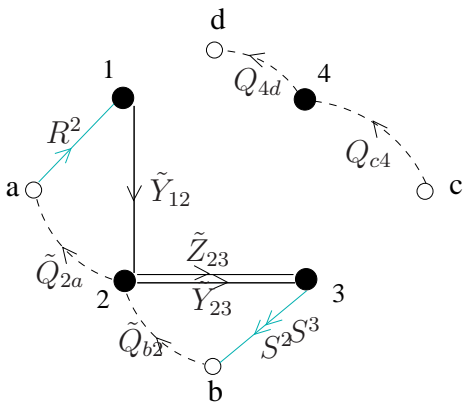}
  \caption{\small Relevant part of dual quiver for second class of bi-fundamentals.}
  \label{dualarbquiv5}
\end{figure}

These fields and their mesons only appear in one term, so will always couple in the same way. Taking as an example
\begin{eqnarray}
  h\, R^2_{a1}\tilde{Y}_{12}\tilde{Q}_{2a} & = &  \pmatrix{R_1 \cr R_2} \tilde{Y}_{12}\, ( \mu + \phi_3 \, ; \, \phi_4)
  \nonumber \\
  & = & \mu R_1\, \tilde{Y}_{12}\, + \, R_1\, \phi_3 \,\tilde{Y}_{12} + R_2 \,\phi_4\, \tilde{Y}_{12}
\end{eqnarray}
This shows that $R_1$ and $\tilde{Y}_{12}$ get tree-level masses and
$R_2$ gets a mass at 1-loop since it couples to the broken
supersymmetry field $\phi_4$.  The only remaining fields are flavors
like $Q_{c4}$, $Q_{4d}$, which do not transform in a gauge group
adjacent to the dualized node (i.e. not adjacent in the quiver loop
corresponding to the superpotential term used to introduce flavors).
These are directly massive from the tree-level $W_{mass}$ term.

So, as stated, all fields except those that appear in the original
superpotential (i.e. mesons with gauge indices and bi-fundamentals
which are not dualized) get masses either at tree-level or at
one-loop. So we only need to check the dualized original
superpotential to see if we have a metastable vacua.

\end{document}